\documentclass[a4paper,11pt]{article}
\pdfoutput=1
\usepackage{jheppub}
\usepackage{graphicx,hyperref,xcolor,amsmath,amssymb,bm,multirow,float,array}
\allowdisplaybreaks
\newcommand{\as}{\alpha_s}
\newcommand{\asc}{\frac{\alpha_s C_F}{2\pi}}
\newcommand{\eps}{\epsilon}
\newcommand{\euv}{\eps_{\mathrm{UV}}}
\newcommand{\eir}{\eps_{\mathrm{IR}}}
\newcommand{\fms}[1]{{#1}\!\!\!/}
\newcommand{\mums}{\mu^{2\eps}_{\overline{\mathrm{MS}}}}
\newcommand{\msbar}{\overline{\text{MS}}}
\newcommand{\Ein}{\mathrm{Ein}}

\newcommand{\Li}{\mathrm{Li}}
\newcommand{\PV}{\mathcal{P}}
\newcommand{\rb}{\bar{r}}
\newcommand{\nb}{\bar{n}}
\newcommand{\bh}{\hat{b}_T}

\begin{document}

\title{Soft factor of quasi-TMDPDFs from the soft limit of the quasi beam function}
\def\KU{Department of Physics, Korea University, Seoul 02841, Korea}
\author[]{Junegone Chay}
\emailAdd{chay@korea.ac.kr}
\affiliation[ ]{\KU}

\abstract{
The quasi-TMDPDF in large-momentum effective theory is related to the TMDPDF through a soft factor. We show how this factor arises from the quasi beam function
itself. Computing the one-loop quasi beam function and decomposing it by the method of regions, we find that the hard and
collinear regions give the matching coefficient and the lightcone beam function. The soft region involves only one lightlike
direction, yet, once the two hemispheres of the gluon momentum are folded onto each other, it reproduces the square root of the lightcone soft function together with the reduced soft function. The
analysis leads to an operator definition of the soft factor, as the vacuum matrix element of the soft Wilson lines into which
the quasi beam function turns. This
definition provides the basis for an all-order analysis, and it shows that the Collins-Soper kernel is reproduced to all
orders.}

\maketitle

\section{Introduction} \label{intro}

Transverse-momentum-dependent parton distribution functions (TMDPDFs) describe the distribution of partons in a hadron
in both the longitudinal momentum fraction and the transverse momentum. They enter the factorization of processes in which a
small transverse momentum is measured, such as Drell-Yan production~\cite{Collins:1984kg} and semi-inclusive deep inelastic scattering~\cite{Ji:2004wu}, and they
carry information on the three-dimensional structure of hadrons that collinear parton distributions do not. Their
definition is more subtle than that of collinear PDFs~\cite{Collins:1981uk,Collins:2011zzd}. A TMDPDF combines a
hadronic matrix element of quark fields joined by Wilson lines, the beam function, with the square root of a vacuum matrix
element of Wilson lines, the soft function. Neither is well defined by itself, and only their combination is, with a
remaining dependence on a rapidity scale that is governed by the Collins-Soper
kernel~\cite{Collins:2011zzd,Becher:2010tm,Echevarria:2011epo,Chiu:2011qc,Chiu:2012ir}.

At small transverse separation $b_T$ the TMDPDFs can be computed in perturbation theory in terms of collinear PDFs, but at
$b_T\sim\Lambda_{\mathrm{QCD}}^{-1}$ they are nonperturbative, and so is the Collins-Soper kernel. A first-principles
determination requires lattice QCD. The Wilson lines of the beam and soft functions are lightlike and extend in time, however,
and such operators cannot be computed directly in Euclidean lattice QCD. Lattice studies of TMDs have therefore used
equal-time operators with staple-shaped Wilson links and ratios in which the soft
factors cancel~\cite{Hagler:2009mb,Musch:2010ka,Musch:2011er,Engelhardt:2015xja,Yoon:2017qzo}.

Large-momentum effective theory (LaMET)~\cite{Ji:2013dva,Ji:2014gla,Ji:2020ect} gets around this obstacle with a
simple observation. A lightcone correlation in a hadron can be viewed as the limit of an equal-time correlation in a hadron moving with
infinite momentum, since a large boost along $\hat{z}$ brings a spacelike separation along $\hat{z}$ close to the lightcone. At
large but finite momentum $P^z$ the equal-time correlation, called a quasi distribution, can be computed in Euclidean lattice
QCD. It has the same infrared, nonperturbative physics as the lightcone distribution, and the two differ only in their
ultraviolet physics, because the limit $P^z\to\infty$ and the ultraviolet regularization are taken in opposite orders. The
difference is perturbative and is absorbed into a matching coefficient, up to corrections suppressed by powers of $1/P^z$. The
simplest example is the quasi-PDF, in which the two quark fields are joined by a straight spacelike Wilson line in place of the
lightlike one of the PDF, and which is related to the PDF by such a perturbative matching coefficient.

For TMDPDFs the quark fields are separated in the transverse direction as well, and a gauge-invariant equal-time operator
requires a Wilson line that runs along $\hat{z}$ to a distance $L$, crosses to the transverse position of the other field, and
comes back~\cite{Ji:2014hxa,Ebert:2018gzl,Ebert:2019okf,Ji:2019sxk,Ji:2019ewn,Vladimirov:2020ofp,Ebert:2022fmh}. The matrix
element of this operator in a hadron, the quasi beam function, inherits the length $L$ of the staple-shaped Wilson line in
several ways. Each segment carries a linear divergence proportional to its length~\cite{Polyakov:1980ca,Dotsenko:1979wb}, the
two antiparallel legs exchange gluons over their whole length and produce a term growing like $L/b_T$, as a static potential
would, and logarithms of $L$ appear as well. The length $L$ is a parameter of the equal-time construction, bounded in lattice
calculations by the size of the lattice, and the dependence on it has to be removed before the quasi beam function can be
related to the TMDPDF.

This is done by dividing by the square root of the vacuum expectation value $Z_E$ of a rectangular
Wilson loop of length $2L$ and width $b_T$. In the spirit of LaMET one would then expect the quotient $\tilde{f}$ to be
related to the TMDPDF by a perturbative matching coefficient. The infrared physics of the two is not the same, however. In the TMDPDF
the soft gluons are described by a soft function, a vacuum matrix element of lightlike Wilson lines, while $Z_E$ is built from
spacelike Wilson lines like the staple itself. At $b_T\sim\Lambda_{\mathrm{QCD}}^{-1}$ soft gluons are nonperturbative, so that
the mismatch cannot be absorbed into the matching coefficient and has to appear as a separate nonperturbative factor. The
relation to the TMDPDF then takes the form~\cite{Ji:2019sxk,Ji:2019ewn}
\begin{equation} \label{factintro}
\tilde{f}(x,b_T,\mu,\zeta_z)\sqrt{S_r(b_T,\mu)} = H\Bigl(\frac{\zeta_z}{\mu^2}\Bigr)\,
e^{\frac{1}{2}K(b_T,\mu)\ln(\zeta_z/\zeta)} f^{\mathrm{TMD}}(x,b_T,\mu,\zeta),
\end{equation}
in which $H$ is the perturbative matching coefficient. The mismatch appears in two pieces. The difference between the
rapidity scales $\zeta_z$ of the quasi-TMDPDF and $\zeta$ of the TMDPDF is governed by the Collins-Soper kernel $K$, and what
remains depends only on $b_T$ and $\mu$ and is called the reduced soft function $S_r$. Relating the
quasi-TMDPDF to the TMDPDF therefore comes down to deciding what the quasi beam function is to be divided by, and the
choice has to be justified by what it does to the relation with the TMDPDF.

For eq.~\eqref{factintro} to be useful, every nonperturbative factor in it has to be computable in lattice QCD, and the
conditions under which lattice calculations are done determine what this requires. A lattice calculation works at finite spacing
$a$, with Wilson lines of finite length $L$ and a hadron of finite momentum $P^z$. The linear divergence of the Wilson line
then appears as a factor that grows with $L/a$, and it is the division by $\sqrt{Z_E}$ that removes it, before the result is
converted from a lattice scheme to the $\msbar$ scheme~\cite{Ebert:2019tvc}. The finite values of $L$ and $P^z$ are tolerable
only because the dependence on $L$ cancels in this way and the corrections in $1/(P^zb_T)$ are small. Within these constraints
lattice QCD has already delivered nonperturbative TMD physics. In the ratio of quasi-TMDPDFs at two momenta the reduced soft
function drops out, and the Collins-Soper kernel has been extracted in this
way~\cite{Ebert:2018gzl,Ebert:2019tvc,Shanahan:2020zxr}. The reduced soft function does not drop out of the quasi-TMDPDF
itself, and it has been obtained from a light-meson form factor~\cite{LatticeParton:2020uhz,Li:2021wvl}. The soft factor is
thus the one ingredient that requires a separate calculation, and how it is defined decides what that calculation has to be.

This soft factor is where the existing approaches differ. Ebert, Stewart and Zhao~\cite{Ebert:2019okf} follow the lightcone
construction and divide by a quasi soft function, a vacuum matrix element of spacelike Wilson lines. The simplest choice, in
which the lines from both quark positions run along $\pm\hat{z}$, leads to logarithms of $b_T$ that do not match those of the
TMDPDF at one loop, and a bent quasi soft function, with lines along two spatial directions, was introduced to restore the
agreement. Ji, Liu and Liu~\cite{Ji:2019sxk,Ji:2019ewn} keep the rectangular loop and attribute the mismatch to $S_r$, which
they define through the soft function with Wilson lines of finite rapidity and extract from the light-meson form factor
mentioned above. Vladimirov and Sch\"afer~\cite{Vladimirov:2020ofp} derived the factorization in soft-collinear effective theory by
representing the staple as the propagator of an auxiliary heavy field, so that the quasi beam function becomes a correlator of
heavy-to-light currents, and Rodini and Vladimirov~\cite{Rodini:2022wki} extended it to next-to-leading power. The
lattice-specific factor of their formula is a vacuum matrix element $\Psi$ whose rapidity evolution is governed by the
Collins-Soper kernel. The factorization formula has since been proved to
all orders in $\alpha_s$ by Ebert, Schindler, Stewart and Zhao~\cite[eq.~(1.1)]{Ebert:2022fmh},
who relate the quasi-TMDPDF through Lorentz invariance to TMDPDFs defined with Wilson lines off the lightcone. In the
approaches of refs.~\cite{Ebert:2019okf,Ji:2019sxk,Ji:2019ewn,Ebert:2022fmh} the soft factor is constructed outside the quasi beam function, since it depends on two almost lightlike
directions and cannot be obtained by boosting an equal-time operator~\cite[section~3.4]{Ebert:2019okf}. An analysis
of the leading regions has been noted as a complement to the proof~\cite[footnote~1]{Ebert:2022fmh}.

The quasi beam function at finite $L$, on the other hand, is a well-defined object, the matrix element of a full-QCD operator
in a hadron. Since the open issue is the soft factor, and soft physics is carried by one region of the loop momenta, it is
natural to decompose this object by the method of regions~\cite{Beneke:1997zp} and to ask whether the quasi beam function
itself determines the soft factor. At leading power in $1/(P^zb_T)$ the loop momenta divide into hard, collinear and soft
regions. We use the lightlike vectors $n^{\mu}=(1,0,0,1)$, along the motion of the hadron, and $\nb^{\mu}=(1,0,0,-1)$, in
terms of which $\hat{z}=(n-\nb)/2$. Collinear gluons see the legs of the Wilson line as a line along $\nb$, so that the collinear region is expected to
reproduce the lightcone beam function once its overlap with the soft region, the zero-bin contribution~\cite{Manohar:2006nz}, is
subtracted. In the soft region the energetic quark supplies a lightlike Wilson line along $n$~\cite{Bauer:2001yt,Chay:2004zn}, while the
legs of the Wilson line along $\hat{z}$, which are long compared with the wavelength of a soft gluon, appear to soft gluons as
soft Wilson lines along $\hat{z}$, in place of the lines along $\nb$ in the soft function of the TMDPDF. In the frame in which the hadron moves fast, the
lightlike direction comes from the state, the energetic quark collinear to $n$, and the spacelike direction from the operator,
which stays at equal time. Viewed from the rest frame of the hadron, the boost brings $\hat{z}$ close to the lightcone along
$\nb$, not along $n$, with an orientation that depends on the process, so that the Wilson line takes the place of the lines
along $\nb$, as the collinear and soft regions confirm in sections~\ref{regions} and \ref{sectors}. The soft limit of the
quasi beam function is therefore expected to be a definite vacuum matrix element, obtained rather than chosen. It contains a
single lightlike direction, whereas the soft function of the TMDPDF contains two, and it is not obvious how the one could
account for the other.

We answer this question at one loop, in three steps, each of which says something about the question. First, to be sure of the
object to be dissected, we compute the quasi beam function at finite $L$ and finite $P^zb_T$ with momentum-space Feynman rules
and reproduce the result of ref.~\cite{Ebert:2019okf}. Every diagram is finite apart from poles in $\eps$, and no rapidity
regulator is needed, since the eikonal lines are spacelike and the hadron carries a finite momentum. This gives an unambiguous
quantity against which every later decomposition can be checked. Second, we decompose the same calculation by the method of
regions. The collinear and soft regions contain rapidity divergences, which we regulate with the hemisphere form of the $\eta$ regulator~\cite{Chiu:2012ir,Chay:2020jzn}. The hard region gives the
matching coefficient $H$, and the collinear region after the zero-bin subtraction gives the lightcone beam function. The matching coefficient $H$ and the lightcone beam function, the two ingredients of eq.~\eqref{factintro} that are not in
dispute, thus appear directly as parts of the quasi beam function. The collinear region carries a rapidity divergence, which
has to cancel against that of the soft region, since the quasi beam function itself has none.

The third step, the computation of the soft region, takes up this puzzle. In the soft function of the TMDPDF a soft gluon is
exchanged between Wilson lines along $n$ and $\nb$, while in the soft limit of the quasi beam function it is exchanged between
the line along $n$ and the leg of the staple along $\hat{z}$. We show that the two are related by what we call the folding of the hemispheres. The eikonal factor of
the leg distinguishes the gluons moving forward, $k_3>0$, from those moving backward, $k_3<0$, only by its sign, and
interchanging $k_+$ and $k_-$ maps one half of the gluon momenta onto the other while reversing that sign. Folding the backward
half onto the forward one, we find that the soft region of the quasi beam function adds up to exactly one hemisphere of the
lightcone soft function, which at one loop is its square root. The single spacelike direction $\hat{z}$ thus does the work of
the second lightlike direction, and the soft physics of the TMDPDF, including its rapidity dependence, is contained in the
quasi beam function itself. What remains of the soft limit is the self-interaction of the
Wilson line. It exceeds that of half of the rectangular loop by a single straight segment, and this segment is the
reduced soft function. The reduced soft function is then not a correction for an imperfect choice of the divisor but a
definite piece of the quasi beam function. This
explains why the simplest quasi soft function of ref.~\cite{Ebert:2019okf}, which removes only half of the loop, fails, and
why the bent one, whose square root is one open staple at one loop, succeeds.

These one-loop results lead to an operator definition of the soft factor. At leading power the quasi beam function factorizes
as $\tilde{B} = H\,B_n\,S_{n\hat{z}}$, where $B_n$ is the lightcone beam function and $S_{n\hat{z}}$, the soft limit of the
quasi beam function, is the vacuum matrix element of the soft Wilson lines into which the quasi beam function turns. The
reduced soft function becomes a ratio of vacuum matrix elements of Wilson lines, and no soft function has to be constructed
separately. Because this definition does not rely on the one-loop calculation, it provides the basis for an all-order
analysis. It shows that the rapidity divergences of the soft limit are one half of those of the lightcone soft function to all
orders, so that the Collins-Soper kernel is reproduced, and it supplies the analysis of the leading regions that complements
the proof of ref.~\cite{Ebert:2022fmh}. For lattice calculations, $S_{n\hat{z}}$ contains a lightlike Wilson line and does not replace
the separate determination of the soft factor. It does support the structure of eq.~\eqref{factintro}, with the
division by $\sqrt{Z_E}$ and the reduced soft function, and it identifies the separately computed $S_r$ as a definite part of
the quasi beam function.

The paper is organized as follows. In section~\ref{quasi} we define the quasi beam function, present the one-loop diagrams
and their matrix elements, and give the result at finite $L$. In section~\ref{regions} we analyze the momentum regions and
the rapidity divergences, and in section~\ref{sectors} we evaluate the collinear and soft sectors with the rapidity regulator,
fold the two hemispheres of the soft region, and show that the sum of the sectors reproduces the full result. The connection
with the factorization formula at one loop is made in section~\ref{softfactor}. In section~\ref{operator} we derive the soft
limit as an operator, express $S_r$ in terms of Wilson lines, and extend the factorization to all orders. We compare with the
quasi soft functions, discuss the role of the staple length and the open questions beyond one loop in
section~\ref{discussion}, and conclude in section~\ref{conclusion}. The Feynman rules for the
staple-shaped Wilson line are derived in appendix~\ref{app:feyn}, and the contributions of the individual diagrams
after the integration over $k^0$ are listed in appendix~\ref{app:k0}.

\section{Quasi beam function at finite staple length} \label{quasi}

\subsection{Definition}

\begin{figure}[t]
\centering
\includegraphics[width=0.85\textwidth]{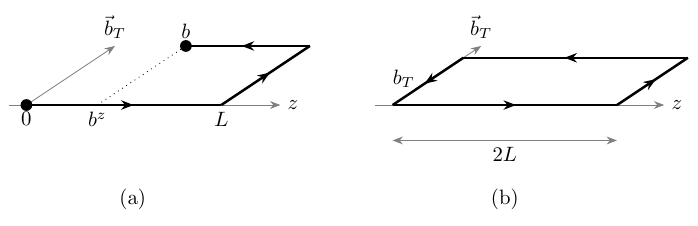}
\caption{\label{figpaths} Wilson lines of the quasi-TMDPDF. (a) The staple-shaped Wilson line of the quasi beam function, with
the quark fields at $0$ and $b=(0,\vec{b}_T,b^z)$. (b) The rectangular Wilson loop of length $2L$ and width $b_T$, whose vacuum
expectation value is $Z_E$.}
\end{figure}

The quasi beam function is the Fourier transform of the spatial correlation of a quark bilocal operator, in which the two quark
fields are separated both longitudinally and transversely and are connected by a staple-shaped Wilson line. In coordinate space
the bare correlation is
\begin{equation} \label{Bcoord}
\tilde{\mathcal{B}} (b^z, \vec{b}_T, P^z, L, \eps) \equiv \langle P| \overline{\psi} (b) \frac{\gamma^z}{2} W_{\sqsupset} (b,0;L)
\psi(0) |P\rangle , \qquad b^{\mu} = (0, \vec{b}_T, b^z),
\end{equation}
and its Fourier transform is
\begin{equation} \label{Bfour}
\tilde{B} (x, \vec{b}_T, P^z, L, \eps) = \int_{-\infty}^{\infty} \frac{db^z}{2\pi} e^{ixP^z b^z}
\tilde{\mathcal{B}} (b^z, \vec{b}_T, P^z, L, \eps).
\end{equation}
The staple $W_{\sqsupset}(b,0;L)$ starts at the origin, runs along $\hat{z}$ to $L\hat{z}$, along $\vec{b}_T$ to
$L\hat{z}+\vec{b}_T$, and returns along $-\hat{z}$ to $b$. Its two longitudinal legs have the lengths $L$ and $L-b^z$,
and each segment is a path-ordered exponential of the form
\begin{equation}
W_{\hat{z}}(\xi;L) = P\exp\left[-ig\int_{\xi^z}^{L} ds\, \hat{z}\cdot A(\vec{\xi}_T + \hat{z}s)\right]
\end{equation}
for the leg between $\xi^z$ and $L$ at the transverse position $\vec{\xi}_T$. The transverse link $W_T(L;\vec{b}_T)$ is the
corresponding straight line at $z=L$ between $\vec{0}_T$ and $\vec{b}_T$. The staple is shown in fig.~\ref{figpaths}(a). In refs.~\cite{Ji:2019sxk,Ji:2019ewn} the quark
fields are placed symmetrically at $\pm b^z/2$~\cite[eq.~(8)]{Ji:2019ewn}, and the two legs have the lengths $L\mp b^z/2$. The two choices differ by a
translation along $z$ and by terms proportional to $|b^z|/b_T$ in the linear divergence of the staple. In the region relevant for
factorization, where $b^z\sim 1/P^z$ is much smaller than $b_T$, these terms are of order $1/(P^z b_T)$ and do not affect
the results at leading power.

The dependence on $L$ is removed by dividing by the square root of the vacuum expectation value of a rectangular
Wilson loop of length $2L$ and width $b_T$, shown in fig.~\ref{figpaths}(b), and the quasi-TMDPDF used in the factorization is
\begin{equation} \label{ftilde}
\tilde{f} (x, \vec{b}_T, \mu, P^z) = \lim_{L\to\infty} \frac{\tilde{B} (x, \vec{b}_T, P^z, L, \mu)}{\sqrt{Z_E (2L, b_T,\mu)}},
\end{equation}
renormalized in the $\msbar$ scheme. We compute $\tilde{B}$ at one loop at finite $L$ and take the limit only at the end, so
that the cancellation of the dependence on $L$ can be followed explicitly. The limit does not require an infinite staple. The
ratio becomes independent of $L$ up to corrections of order $b_T/L$, and $L\gg b_T$ suffices.

The factor $\sqrt{Z_E}$ removes the dependence on the Wilson line that does not involve the hadron. For long sides the
dependence of a Wilson line on its length exponentiates. The two longitudinal legs, separated by $b_T$, are then like the
world lines of a static quark and antiquark, and both the staple and the loop depend on $L$ as such a pair does. Two kinds of contributions enter. The self-energy of the lines is local and is
proportional to the total length of the contour. With a cutoff regulator it contains a linear divergence, which in lattice
QCD is removed by the Wilson-line mass renormalization~\cite{Polyakov:1980ca,Dotsenko:1979wb,Ji:2017oey,Ishikawa:2017faj,Green:2017xeu}. The exchange between the antiparallel lines is proportional to the
length over which they run side by side, and it produces the static potential~\cite{Ji:2019sxk,Ji:2019ewn}. At leading power, where $b^z$ can be set to zero as discussed above, the staple has the total length $2L+b_T$ and
antiparallel legs of length $L$, and half of the loop with sides $2L$ and $b_T$ has the same total length and the same length
of antiparallel sides. Both contributions therefore cancel in the ratio to all orders, up to corrections of order $b_T/L$. In
dimensional regularization the linear divergence is absent, and what remains visible is the potential, which at one loop
gives the term $2\pi L/b_T$.
The division by $\sqrt{Z_E}$, however, does not remove every divergence of the Wilson line. The staple ends on the quark fields,
whereas a closed loop has no ends. In the sail diagram of fig.~\ref{figdiag}(a) and its mirror image, a gluon from the Wilson line is attached
to an external quark line, and these diagrams contain an ultraviolet divergence that has no counterpart in the loop. It is
logarithmic and independent of $L$, and it is removed by the $\msbar$ renormalization of the quark bilocal
operator.

The name quasi beam function follows ref.~\cite{Ebert:2019okf}, by analogy with the beam function of SCET, but it should
not be read in the SCET sense. In SCET a beam function is the matrix element of an operator built from collinear fields, and it
describes the collinear modes alone once the soft contribution has been factored out. The quasi beam function in
eq.~\eqref{Bcoord} is instead the matrix element of a full-QCD operator. In the regime of factorization its loop corrections
receive contributions from hard, collinear and soft momenta, and nothing has been factored out of it. Its hard part gives the matching coefficient, its collinear part at
leading power in $1/(P^zb_T)$ is the lightcone beam function in the SCET sense, and its soft part involves the staple-shaped Wilson line and a lightlike one, as shown in section~\ref{regions}. What this soft part contains is the question addressed in this paper.

We use an on-shell massless quark with momentum $p^{\mu}=(p^z,0,0,p^z)$ as the external state, and write $p=p^z$
where no confusion arises. With the average over spin and color, $\frac{1}{2}\sum_s\bar{u}_s(p)\frac{\gamma^z}{2}u_s(p) = p$, and eq.~\eqref{Bcoord} gives
$\tilde{\mathcal{B}}^{(0)} = p\,e^{-ipb^z}$ at tree level, so that eq.~\eqref{Bfour} gives $\tilde{B}^{(0)} = \delta(1-x)$. Both ultraviolet and infrared divergences are regulated in dimensional regularization with
$D=4-2\eps$, and we work in the Feynman gauge. The lightcone components are $k_{\pm}=k^0\pm k_3$, so that
$p\cdot k = p\, k_-$. Throughout the paper we use the abbreviations
\begin{equation} \label{logs}
L_b = \ln \frac{\mu^2 b_T^2}{b_0^2}, \qquad L_z = \ln \frac{(2p^z)^2}{\mu^2}, \qquad
\ell_b = \ln \frac{2p^z b_T}{b_0} = \frac{L_b}{2} + \ln\frac{2p^z}{\mu},
\end{equation}
with $b_0 = 2e^{-\gamma_E}$, and write $b=b_T$ in the arguments of functions.

\subsection{One-loop diagrams}

\begin{figure}[t]
\centering
\includegraphics[width=\textwidth]{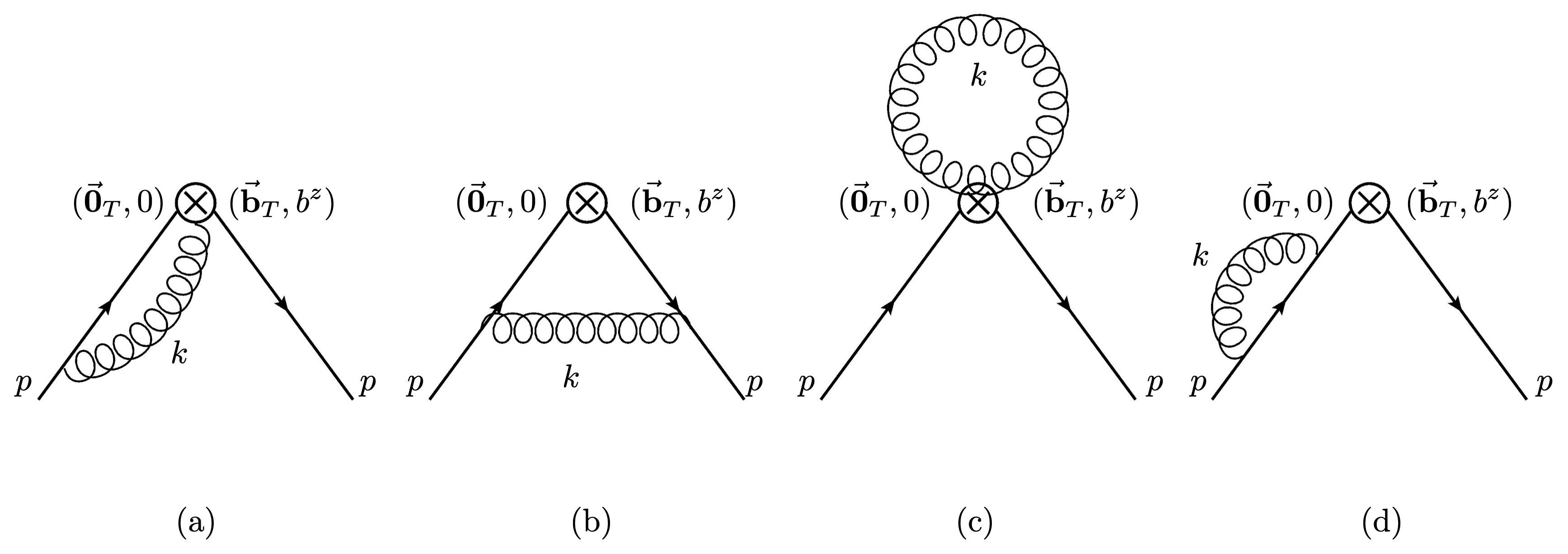}
\caption{\label{figdiag} One-loop diagrams for the quasi beam function in momentum space. The operator vertex $\otimes$ 
represents $\overline{\psi}(b)\,W_{\sqsupset}(b,0;L)\,\psi(0)$, with the quark fields at $0$ and $b$ joined by the Wilson
line shown in fig.~\ref{figpaths}(a). It carries zero, one or two gluons according to the Feynman rules of
appendix~\ref{app:feyn}. (a) Sail diagram, (b) vertex diagram, (c) Wilson-line diagram, which contains all attachments of
the two gluons to the Wilson line, and (d) quark wave-function renormalization. The complex conjugates of (a) and (d) are
omitted.}
\end{figure}

The one-loop diagrams are shown in fig.~\ref{figdiag}. The Wilson line is not drawn explicitly. It is represented by the
operator vertex $\otimes$, which carries no gluon, one gluon or two gluons according to the Feynman rules derived in
appendix~\ref{app:feyn}. Each of these vertices sums over all points of attachment on the Wilson line, namely the two
longitudinal legs and the transverse link, and the positions of the segments enter only through the phases $e^{ik_3L}$,
$e^{ik_3b^z}$ and $e^{i\vec{k}_T\cdot\vec{b}_T}$. The diagrams correspond one to one to those of ref.~\cite{Ebert:2019okf},
where the Wilson line is drawn in coordinate space, and we use the same names. In the sail diagram in fig.~\ref{figdiag}(a),
the gluon from the one-gluon vertex is attached to the incoming quark line. Its mirror image, in which the gluon is attached
to the outgoing quark line, is the complex conjugate and is not drawn. In the vertex diagram in fig.~\ref{figdiag}(b), the gluon is exchanged between the two quark lines
and the operator vertex is the tree-level one. In the Wilson-line diagram in fig.~\ref{figdiag}(c), the two gluons of the
two-gluon vertex are joined by a propagator. It contains the self-energies of the three segments of the Wilson line and the
exchange between its two longitudinal legs. Together they correspond to the Wilson-line self-energy, or tadpole, diagram
of ref.~\cite{Ebert:2019okf}. The
quark wave-function renormalization in fig.~\ref{figdiag}(d) is scaleless, since the on-shell massless quark with
$p^2=0$ provides no dimensionful quantity. In dimensional regularization it vanishes because its ultraviolet and infrared
poles cancel. When the two kinds of poles are distinguished, it converts the infrared poles of the other diagrams into
ultraviolet ones, and it is used in this way in section~\ref{sumdiag}. The Feynman rules are given in
appendix~\ref{app:feyn}. We use $\rb^{\mu}=(0,\vec{0}_T,-1)$, so that $\rb\cdot k = k_3$, the unit vector $\bh$ along 
$\vec{b}_T$, and the abbreviations
\begin{equation}
D_g = k^2+i0, \qquad D_q = (p-k)^2+i0, \qquad k_b = \bh\cdot\vec{k}_T.
\end{equation}
Here $D_g$ and $D_q$ are the denominators of the gluon propagator and of the quark propagator with momentum $p-k$ in the
sail and vertex diagrams of figs.~\ref{figdiag}(a) and (b).

We write the one-gluon vertex in eq.~\eqref{onegluon} as $gT^a\bigl[\rb^{\mu}F_L(k) + \bh^{\mu}F_T(k)\bigr]$ with
\begin{equation}
F_L(k) = \frac{1-e^{ik_3L}+e^{i\vec{k}_T\cdot\vec{b}_T}\bigl(e^{ik_3L}-e^{ik_3b^z}\bigr)}{k_3-i0}, \qquad
F_T(k) = \frac{e^{ik_3L}\bigl(1-e^{i\vec{k}_T\cdot\vec{b}_T}\bigr)}{k_b-i0}.
\end{equation}
Each term records the point at which the gluon is attached. In $F_L$, the term $1-e^{ik_3L}$ comes from the leg at the
origin, which runs from $0$ to $L$. The term $e^{i\vec{k}_T\cdot\vec{b}_T}e^{ik_3L}$ comes from the far end of the leg at
$\vec{b}_T$, and $e^{i\vec{k}_T\cdot\vec{b}_T}e^{ik_3b^z}$ from its near end, where it meets the quark field at $b$. $F_T$
is the transverse link at $z=L$. Only the last term of $F_L$ knows about $b^z$, and this will decide which contributions fix
the momentum of the gluon.
The sail diagram in fig.~\ref{figdiag}(a) is then
\begin{align} \label{Msail}
\tilde{B}^{(1)}_{\mathrm{(a)}} &= ig^2C_F\,\mums\int_{-\infty}^{\infty}\frac{db^z}{2\pi}\, e^{-ipb^z(1-x)}
\int\frac{d^Dk}{(2\pi)^D}\, \frac{1}{D_gD_q} \nonumber \\
&\quad\times\frac{1}{2}\sum_s\bar{u}_s(p)\,\frac{\gamma^z}{2}\,(\fms{p}-\fms{k})
\bigl[\fms{\rb}\,F_L(k) + \fms{\hat{b}}_T\,F_T(k)\bigr]u_s(p) \nonumber \\
&= ig^2C_F\,\mums\, p\int_{-\infty}^{\infty}\frac{db^z}{2\pi}\, e^{-ipb^z(1-x)}
\int\frac{d^Dk}{(2\pi)^D}\, \frac{1}{D_gD_q} \nonumber \\
&\quad\times\Bigl\{ \bigl[2(p-k_3)-k_-\bigr]F_L(k) + k_b\,F_T(k)\Bigr\}.
\end{align}
Its mirror image is the complex conjugate of eq.~\eqref{Msail}. The vertex diagram in fig.~\ref{figdiag}(b) is
\begin{align} \label{Mvertex}
\tilde{B}^{(1)}_{\mathrm{(b)}} &= -ig^2C_F\,\mums\int\frac{d^Dk}{(2\pi)^D}\,\delta\bigl(k_3-p(1-x)\bigr)\,
\frac{e^{i\vec{k}_T\cdot\vec{b}_T}}{D_gD_q^2} \nonumber \\
&\quad\times\frac{1}{2}\sum_s\bar{u}_s(p)\,\gamma_{\mu}(\fms{p}-\fms{k})\,
\frac{\gamma^z}{2}\,(\fms{p}-\fms{k})\gamma^{\mu}u_s(p) \nonumber \\
&= -ig^2C_F\,\mums\,(2-D)\int\frac{d^Dk}{(2\pi)^D}\,\delta\bigl(k_3-p(1-x)\bigr)\,
\frac{e^{i\vec{k}_T\cdot\vec{b}_T}}{D_gD_q^2}\bigl[-k_3(p-k)^2-(p-k_3)k^2\bigr] \nonumber \\
&= ig^2C_F\,\mums\,(2-D)\int\frac{d^Dk}{(2\pi)^D}\,\delta\bigl(k_3-p(1-x)\bigr)\,
e^{i\vec{k}_T\cdot\vec{b}_T}\left[ \frac{k_3}{D_gD_q} + \frac{p-k_3}{D_q^2}\right].
\end{align}
With the two-gluon vertex in eq.~\eqref{twogluon}, the Wilson-line diagram in fig.~\ref{figdiag}(c) is
\begin{align} \label{MWilson}
\tilde{B}^{(1)}_{\mathrm{(c)}} &= \frac{1}{2}\,g^2C_F\,\mums\int_{-\infty}^{\infty}\frac{db^z}{2\pi}\, e^{-ipb^z(1-x)}
\int\frac{d^Dk}{(2\pi)^D}\,\frac{-ig_{\mu\nu}}{D_g}\bigl[A^{\mu\nu}(k)+A^{\mu\nu *}(k)\bigr] \nonumber \\
&\quad\times\frac{1}{2}\sum_s\bar{u}_s(p)\,\frac{\gamma^z}{2}\,u_s(p) \nonumber \\
&= \frac{1}{2}\,g^2C_F\,\mums\, p\int_{-\infty}^{\infty}\frac{db^z}{2\pi}\, e^{-ipb^z(1-x)}
\int\frac{d^Dk}{(2\pi)^D}\,\frac{i}{D_g}\Bigl[\mathcal{A}(k)+\mathcal{A}^*(k)\Bigr],
\end{align}
with $\mathcal{A}(k) = -g_{\mu\nu}A^{\mu\nu}(k)$. Since $-\rb^2=-\bh^2=1$, $\mathcal{A}(k)$ is the sum of the coefficients of
$\rb^{\mu}\rb^{\nu}$ and $\bh^{\mu}\bh^{\nu}$ in $A^{\mu\nu}$, that is, of the contributions of the longitudinal legs and of
the transverse link given in appendix~\ref{app:feyn}. The factor $1/2$ is the symmetry factor of the two-gluon vertex.

The integration over $b^z$ separates every diagram into two types of contributions. A term which does not depend on $b^z$
gives
\begin{equation} \label{virtdef}
\int_{-\infty}^{\infty}\frac{db^z}{2\pi}\, e^{-ipb^z(1-x)} = \frac{1}{p}\,\delta(1-x),
\end{equation}
and we call it virtual. A term with the phase $e^{ik_3b^z}$ gives
\begin{equation} \label{realdef}
\int_{-\infty}^{\infty}\frac{db^z}{2\pi}\, e^{-ipb^z(1-x)}e^{ik_3b^z} = \delta\bigl(k_3-p(1-x)\bigr),
\end{equation}
which fixes the longitudinal momentum of the gluon, and we call it real. The quark field at $b$ then carries the momentum
fraction $x$, which is not restricted to $0<x<1$ since $k_3$ takes either sign. The distinction refers to the momentum flow
through the operator and not to the emission of an on-shell gluon. The quasi beam function is the matrix element of an
equal-time operator, not a squared amplitude, and has no final state into which a gluon could be emitted. The phase
$e^{ik_3b^z}$ arises only when the gluon is attached at a point whose position is tied to the quark field at $b$, and this
decides for every term whether it is virtual or real. Table~\ref{tabcontrib} follows each point of attachment through the
calculation, from the term in the Feynman rule to its label in appendix~\ref{app:k0}, where the contributions are listed after
the integration over $k^0$, and to the result in section~\ref{secresults}. The two-gluon vertex also contains terms proportional to the lengths of the segments, among them a term linear
in $b^z$ from the leg of length $L-b^z$, whose Fourier transform would give a derivative of $\delta(1-x)$. With the length
factored out, however, the remaining integral over $k$ contains neither the phase $e^{i\vec{k}_T\cdot\vec{b}_T}$ nor any other
scale. These terms are the linear divergence of the Wilson line, which is scaleless and vanishes in dimensional
regularization.

\begin{table}[t]
\centering
\small
\renewcommand{\arraystretch}{1.2}
\begin{tabular}{llccc}
\hline
Diagram & Attachment & Type & Appendix~\ref{app:k0} & Section~\ref{secresults} \\
\hline
Sail & leg at $\vec{0}_T$ & V & $V^a_1$--$V^a_4$ & \multirow{2}{*}{sail, v1 and v2} \\
 & far end of leg at $\vec{b}_T$ & V & $V^a_7$--$V^a_{10}$ & \\
 & transverse link & V & $V^a_5$, $V^a_6$ & sail, v2 \\
 & near end of leg at $\vec{b}_T$ & R & $R^a_1$--$R^a_4$ & sail, r \\
\hline
Vertex & -- & R & $R^b_1$--$R^b_3$ & vertex \\
\hline
Wilson line & legs, independent of $b^z$ & V & $V^c_1$ & leg \\
 & leg of length $L-b^z$ & R & $R^c_1$ & leg$'$ \\
 & transverse link & V & $V^c_T$ & link \\
 & exchange between the legs & V and R & $V^c_3$, $R^c_2$ & exch \\
 & terms linear in the lengths & V & $V^c_2$ & scaleless \\
\hline
\end{tabular}
\caption{\label{tabcontrib} The points of attachment of the gluons to the Wilson line, whether they give virtual (V) or
real (R) contributions, their labels after the integration over $k^0$ in appendix~\ref{app:k0}, and the subscripts with which the
corresponding results are labeled in section~\ref{secresults}. For the Wilson-line diagram the first three rows are self-energies. In the sail diagram the terms from the leg at $\vec{0}_T$ and from the far end of the leg
at $\vec{b}_T$ are regrouped by the identity in eq.~\eqref{splitid} into a part depending only on $p$ (v1) and a part
depending on $L$ and $b_T$ (v2).}
\end{table}

\subsection{Evaluation} \label{presc}

Every contribution is evaluated in the same three steps, the integration over $k^0$, the transverse integration at fixed
$k_3$, and the treatment of the remaining dependence on $k_3$, or on $x$, as a distribution. The order is chosen so that each
step produces a definite kind of singularity. The integration over $k^0$ produces none, all poles in $\eps$ arise in the
transverse integration, and the last step deals only with the singular points in $k_3$. Each diagram is well defined, but
some of the terms produced along the way are not, and in each step we indicate how the terms have to be grouped.

\paragraph{Integration over $k^0$.} Closing the contour in $k^0$ splits each term of the Feynman rules into the residue at the
gluon pole and that at the quark pole. For the sail diagram the sum of the two residues is
\begin{equation} \label{k0sail}
\int dk^0\, \frac{2(p-k_3) -k_-}{(k^2 +i0) [(p-k)^2 +i0]} = \frac{i\pi}{p}\, h(k_3,k_T), \qquad
h = h_1+h_2+h_3+h_4,
\end{equation}
with
\begin{equation} \label{hj}
h_1 = \frac{(p-k_3)\,k_3}{k_T^2\,E}, \qquad h_2 = \frac{(p-k_3)^2}{k_T^2\,E_q}, \qquad
h_3 = -\frac{1}{2E}, \qquad h_4 = \frac{1}{2E_q},
\end{equation}
where $E = (k_3^2 + k_T^2)^{1/2}$ and $E_q = [(p-k_3)^2 + k_T^2]^{1/2}$. For $k_3<0$ each residue grows linearly in $|k_3|$ and is singular as $1/k_T^2$, although the gluon moving
backward cannot be collinear to the quark. These terms cancel in the sum, which falls off as $1/k_3^2$ for $|k_3|\to\infty$
and is singular at $k_T\to 0$ only for $0<k_3<p$, where the gluon is collinear to the quark. The two residues are therefore
never integrated separately, and the terms $h_j$ in eq.~\eqref{hj} are a decomposition of their sum that does not correspond to the individual residues.\footnote{The
residue at the gluon pole is $-p(E-k_3+2p)/[E(E-E_q+p)(E+E_q+p)]$ and that at the quark pole is
$p(E_q-k_3+p)/[E_q(E-E_q+p)(E+E_q-p)]$, in units of $i\pi/p$. For $k_3<0$ the factor $E-E_q+p$ vanishes as $k_T\to 0$.
For $0<k_3<p$ the bracket in $h_1+h_2$ approaches $\mathrm{sgn}(k_3)+\mathrm{sgn}(p-k_3) = 2$ as $k_T\to 0$.} The
expansion of $h$ in section~\ref{regions} defines the hard, collinear and soft regions. For the vertex diagram the corresponding combination is
\begin{align} \label{k0vertex}
&\int dk^0\left[ \frac{k_3}{(k^2+i0)[(p-k)^2+i0]} + \frac{p-k_3}{[(p-k)^2+i0]^2}\right] \nonumber \\
&\qquad = \frac{i\pi}{2}\left[\frac{k_3^2}{pk_T^2 E} + \frac{k_3(p-k_3)}{pk_T^2E_q} +\frac{p-k_3}{E_q^3}\right],
\end{align}
and for the Wilson-line diagram only the gluon propagator is involved, $\int dk^0/(k^2+i0) = -i\pi/E$. In
appendix~\ref{app:k0} the four terms $h_j$ give the contributions $V^a_j$ and $R^a_j$ of the sail diagrams, the three terms
of eq.~\eqref{k0vertex} give $R^b_1$, $R^b_2$ and $R^b_3$ of the vertex diagram with $k_3=p(1-x)$, and the factor $1/E$
appears in all the contributions of the Wilson-line diagram.

\paragraph{Transverse integration.} At fixed $k_3$ the integrands depend on $\vec{k}_T$ through $k_T^2$, through the phase
$e^{i\vec{k}_T\cdot\vec{b}_T}$, and in the transverse link through $k_b$. Where the integrand behaves as $1/k_T^2$ at small
$k_T$, we split the phase as $e^{i\vec{k}_T\cdot\vec{b}_T} = 1+\bigl(e^{i\vec{k}_T\cdot\vec{b}_T}-1\bigr)$. The term without
the phase carries the collinear divergence and is evaluated in $2-2\eps$ dimensions, while the term with
$e^{i\vec{k}_T\cdot\vec{b}_T}-1$ is finite and is evaluated at $\eps=0$. All poles in $\eps$ arise in this step, through the
functions $G(\eps)$ and $H(\eps)$ defined below, and the finite Fourier integrals produce the functions $\Ein$, $E_1$ and
exponentials of $p|x|b_T$ and $p|1-x|b_T$. The terms linear in $L$, $b^z$ and $b_T$ in the two-gluon vertex lead to transverse
integrals without a scale. They are the linear divergence of the Wilson line, which would appear with a cutoff regulator, and
they vanish in dimensional regularization.

\paragraph{Dependence on $k_3$.} The eikonal propagator of the longitudinal leg carries the prescription $1/(k_3-i0)$ in the
sail diagram of fig.~\ref{figdiag}(a), and $1/(k_3+i0)$ in its complex conjugate. In the real contributions, where
$k_3 = p(1-x)$, the sum of the two produces the principal value of $1/(1-x)$. In the virtual contributions, the leg at
the origin appears through the combination $(1-e^{ik_3L})/k_3$, which is regular at $k_3=0$, while the term
$e^{ik_3L} e^{i\vec{k}_T\cdot\vec{b}_T}/(k_3-i0)$ from the leg at $\vec{b}_T$ keeps its prescription. Since the sum of a
diagram and its mirror image is real, only the real part of each contribution is retained. In the Wilson-line diagram the two
eikonal propagators of the same leg appear as the pinched product $1/[(k_3-i0)(k_3+i0)]$, which causes no difficulty. In the
exchange between the legs it multiplies $1-e^{ik_3L}$, which vanishes at $k_3=0$, and in the self-energies the transverse
integration produces $|k_3|^{1-2\eps}$, which softens the singularity to $|k_3|^{-1-2\eps}$. Factors such as
$|1-x|^{-1-2\eps}$ are then expanded as distributions in $x$, as described in section~\ref{sumdiag}.

\subsection{Results for the individual diagrams} \label{secresults}

We present the contributions of the diagrams in groups, each including the mirror image where it applies. We use
\begin{equation}
K_{\eps} = \frac{(\mu^2 e^{\gamma_E})^{\eps}}{\Gamma(1-\eps)}, \qquad
G(\eps) = \frac{\Gamma(-\eps)\Gamma(\frac{1}{2}+\eps)}{\sqrt{\pi}}, \qquad
H(\eps) = \frac{\Gamma(1-\eps)\Gamma(\eps-\frac{1}{2})}{\sqrt{\pi}},
\end{equation}
and the function
\begin{equation}
\Ein(u) = \int_0^u dt\, \frac{1-e^{-t}}{t} = \gamma_E + \ln u + E_1(u),
\end{equation}
where
\begin{equation}
E_1(u) = \int_u^{\infty}dt\,\frac{e^{-t}}{t}
\end{equation}
is the exponential integral. The function $\Ein(u)$ is regular at $u=0$, where $\Ein(u)\simeq u$, and the second form
separates its behavior at large $u$. The logarithm $\ln u$ carries the dependence on $p^zb_T$ that survives for $p^zb_T\gg 1$,
while $E_1(u)\simeq e^{-u}/u$ is exponentially small for large $u$. With $\Ein$ we define
\begin{equation}
\mathcal{E}(x) = \mathrm{sgn}(1-x)\, \Ein\bigl(p|1-x|b\bigr) + \mathrm{sgn}(x)\, \Ein\bigl(p|x|b\bigr).
\end{equation}
The function $\mathcal{E}(x)$ arises from the transverse Fourier integral of the collinear propagator with $e^{i\vec{k}_T\cdot
\vec{b}_T}-1$, and the dimensionally regulated part is the integral without the phase.

\paragraph{Virtual sail.} The virtual contribution of the sail diagrams can be reorganized using the identity
\begin{equation} \label{splitid}
\frac{(1-e^{ik_3L}) + e^{ik_3L}e^{i\vec{k}_T\cdot\vec{b}_T}}{k_3-i0} = \frac{1}{k_3-i0} +
\frac{e^{ik_3L}(e^{i\vec{k}_T\cdot\vec{b}_T}-1)}{k_3-i0}.
\end{equation}
The first term depends neither on $b_T$ nor on $L$, and its contribution depends on $p$ alone. It is
\begin{align} \label{sailh}
\tilde{B}^{(1)}_{\mathrm{sail,v1}} &= \asc\, \delta(1-x)\, K_{\eps}\, p^{-2\eps}\, \pi\cot(\pi\eps)
\Bigl[ G(\eps)+\frac{1}{2} H(\eps) \Bigr] \nonumber \\
&= \asc\, \delta(1-x) \left[ -\frac{1}{\eps^2} - \frac{1}{\eps}\Bigl(1+2\ln\frac{\mu}{2p}\Bigr) - 2\ln^2\frac{\mu}{2p}
-2\ln\frac{\mu}{2p} - 2 + \frac{\pi^2}{12} \right].
\end{align}
It has no counterpart in the lightcone beam function, and in section~\ref{hardreg} it is identified with the hard region.
The second term in eq.~\eqref{splitid}, together with the attachment to the transverse link, gives
\begin{align}
\tilde{B}^{(1)}_{\mathrm{sail,v2}} = \asc\, \delta(1-x) &\left\{ -\frac{\sin pL}{pL} \ln\Bigl(1+\frac{b^2}{L^2}\Bigr) -
\Li_2\Bigl(-\frac{b^2}{L^2}\Bigr) \right. \nonumber \\
&\left. + \frac{2}{p}\int_L^{\infty} d\lambda\, \sin(p\lambda) \left[ \frac{1}{\lambda^2+b^2} - \frac{1}{\lambda^2}
-\frac{1}{\lambda^2}\ln\Bigl(1+\frac{b^2}{\lambda^2}\Bigr)\right] \right\},
\end{align}
which is finite and vanishes as $b^2/L^2$ or $1/(pL)^2$ for large $L$. The reason can be seen in the second term of eq.~\eqref{splitid}. Its phase $e^{ik_3L}$ oscillates rapidly
unless $k_3\lesssim 1/L$, and its factor $e^{i\vec{k}_T\cdot\vec{b}_T}-1$ vanishes unless $k_T\gtrsim 1/b_T$. The term is
therefore confined to $k_3\lesssim 1/L\ll 1/b_T\lesssim k_T$, a corner of momentum space that shrinks as $L$ grows, and it
does not contribute at leading power.

\paragraph{Real sail.} With $k_3=p(1-x)$, the real contribution of the sail diagrams is
\begin{align} \label{sailr}
\tilde{B}^{(1)}_{\mathrm{sail,r}} = \asc &\left\{ K_{\eps}\, G(\eps)\, p^{-2\eps}\, \frac{x}{1-x} \Bigl[ \mathrm{sgn}(1-x)
|1-x|^{-2\eps} + \mathrm{sgn}(x) |x|^{-2\eps} \Bigr] - \frac{2x}{1-x} \mathcal{E}(x) \right. \nonumber \\
&\left. + \frac{e^{-p|x|b} - e^{-p|1-x|b}}{pb(1-x)} \right\},
\end{align}
where $1/(1-x)$ is the principal value. The first term contains two divergences. For $0<x<1$ the factor $G(\eps)$ carries the
collinear divergence, from gluons moving along the quark. The first term in the bracket also combines with the prefactor into
$x|1-x|^{-1-2\eps}$, which is not integrable at $x=1$ for $\eps=0$. There $k_3=p(1-x)$ vanishes, and since the factor
$|1-x|^{-2\eps}$ shows that the transverse momentum is of order $|k_3|$, the whole momentum of the gluon becomes soft.
Dimensional regularization makes the factor integrable, and its expansion as a distribution in section~\ref{sumdiag} turns
the soft divergence into a pole multiplying $\delta(1-x)$. The real sail is the only diagram that contributes to both the collinear
and the soft regions of section~\ref{regions}, and the soft factor emerges from the expansion of its integrand.

\paragraph{Vertex diagram.} The vertex diagram has only a real contribution, since the gluon is attached to both quark lines
and always carries the phase $e^{ik_3b^z}$,
\begin{align}
\tilde{B}^{(1)}_{\mathrm{vertex}} = \asc &\left\{ \frac{1-\eps}{2}\, K_{\eps}\, G(\eps)\, p^{-2\eps} \Bigl[ |1-x|^{1-2\eps} +
(1-x)\,\mathrm{sgn}(x)\, |x|^{-2\eps} \Bigr] \right. \nonumber \\
&\left. - (1-x)\, \mathcal{E}(x) + \mathrm{sgn}(x)\, e^{-p|x|b} \right\},
\end{align}
and it is regular at $x=1$. For $p^zb_T\gg 1$ it reduces to the vertex diagram of the lightcone beam function, and it
belongs to the collinear sector.

\paragraph{Wilson-line diagrams.} The self-energies of the segments of the Wilson line have a universal form. For a straight
spacelike Wilson line of length $\ell$, the self-energy is
\begin{equation} \label{selfen}
\asc\,\Sigma(\ell), \qquad \Sigma(\ell) = \frac{1}{\eps} + 2\ln \frac{\mu \ell}{b_0} + 2,
\end{equation}
independent of the orientation of the line, since after the integration over $k^0$ the integrand depends only on the spatial
momentum and on its component along the line. For the transverse link the component along the line is $k_b$, and its
contribution has the form of that of a longitudinal leg with $k_3\to k_b$ and $L\to b_T$. The contribution of the transverse
link is independent of $p^z$, but it enters the coefficient of the $\delta(1-x)$ pole and that of $L_b$. The same
self-energy appears in the lightcone soft function regulated by Wilson lines of finite length, whose two transverse links
each contribute $\asc[1/\eps + L_b + 2]$~\cite[fig.~4(d) and eq.~(3.16)]{Ebert:2019okf}. The self-energy of
the longitudinal leg of length $L$ and that of the transverse link are then
\begin{align} \label{leglink}
\tilde{B}^{(1)}_{\mathrm{leg}} &= \asc\, \delta(1-x)\, \frac{(\mu^2 e^{\gamma_E}L^2)^{\eps}}{\Gamma(1-\eps)} H(\eps)
\Gamma(-2\eps)\cos(\pi\eps) \nonumber \\
&= \asc\, \delta(1-x) \left[ \frac{1}{\eps} + 2\ln\frac{\mu L}{b_0} + 2 \right], \nonumber \\
\tilde{B}^{(1)}_{\mathrm{link}} &= \asc\, \delta(1-x) \left[\frac{1}{\eps} + L_b + 2\right].
\end{align}
The self-energy of the leg of length $L-b^z$ depends on $b^z$ and therefore gives a real contribution,
\begin{equation}
\tilde{B}^{(1)}_{\mathrm{leg}'} = \frac{1}{2}\asc\, K_{\eps}\, H(\eps)\, p^{-2\eps}\, |1-x|^{-1-2\eps} \cos\bigl(p(1-x)L\bigr).
\end{equation}
The exchange between the two legs contributes
\begin{align}
\tilde{B}^{(1)}_{\mathrm{exch}} = \asc &\left\{ \delta(1-x) \left[ \frac{2L}{b} \arctan\frac{L}{b}
- \ln\Bigl(1+\frac{L^2}{b^2}\Bigr)\right] \right. \nonumber \\
&\left. + \frac{\bigl[1-\cos\bigl(p(1-x)L\bigr)\bigr]\, e^{-p|1-x|b}}{pb(1-x)^2} \right\}.
\end{align}
The integral over $x$ of the real contribution of the exchange is equal to the coefficient of $\delta(1-x)$ in its virtual
contribution.
The Wilson-line diagrams do not involve the quark, and in section~\ref{sectors} they become the part $\Delta_W$ of the soft
sector.

\subsection{Sum of the diagrams} \label{sumdiag}

To combine the contributions, we expand the singular factors in $\eps$ as distributions. With the notation for distributions on the full line introduced in
ref.~\cite{Chay:2026mft}, we have
\begin{align} \label{distexp}
|1-x|^{-1-2\eps} = {}&-\frac{1}{\eps}\delta(1-x) + \Bigl[ \frac{1}{|1-x|} \Bigr]_{+(1)}^{[0,2]} + \frac{\theta(|1-x|>1)}{|1-x|}
\nonumber \\
&- 2\eps \left\{ \Bigl[ \frac{\ln|1-x|}{|1-x|} \Bigr]_{+(1)}^{[0,2]} + \frac{\theta(|1-x|>1)\ln|1-x|}{|1-x|}\right\} +
\mathcal{O}(\eps^2),
\end{align}
and we denote the distributions in the $\mathcal{O}(\eps^0)$ and $\mathcal{O}(\eps)$ terms by $\mathcal{D}_0(x)$ and
$\mathcal{D}_1(x)$. The expansion holds on the whole line. The plus prescription acts only near the singular point, on the
interval $|1-x|<1$, which is symmetric about $x=1$, and the region $|1-x|>1$ is kept as an ordinary function. The unit
half-width makes the coefficient of $\delta(1-x)$ exactly $-1/\eps$. The cancellations at $x=1$ can be seen before the expansion in $\eps$. The first term in
eq.~\eqref{distexp} gives the real sail the $\delta(1-x)$ part $-\eps^{-1}K_{\eps}G(\eps)p^{-2\eps}$, while the virtual sail in
eq.~\eqref{sailh} contains the same factor $K_{\eps}G(\eps)p^{-2\eps}$ multiplied by $\pi\cot(\pi\eps)$. Since
$\pi\cot(\pi\eps)=1/\eps-\pi^2\eps/3+\mathcal{O}(\eps^3)$, the two cancel up to $\pi^2/3$, and with them the double pole and
the single pole proportional to $\ln(\mu/2p)$. In the same way the term with $H(\eps)$ in the virtual sail cancels against
the $\delta(1-x)$ part of $\tilde{B}^{(1)}_{\mathrm{leg}'}$, which carries the same factor $\frac{1}{2}K_{\eps}H(\eps)p^{-2\eps}$,
and nothing of it remains as $\eps\to 0$.
Adding all contributions, we obtain
\begin{align} \label{result}
\tilde{B}^{(1)} = \asc &\left\{ -\frac{1}{\eps} \Bigl[ P_{qq}(x) - \frac{7}{2}\delta(1-x)\Bigr] + \mathcal{D}\,\delta(1-x)
\right. \nonumber \\
&+ \left[ 2\ln\frac{2p}{\mu} \Bigl( 2x \Bigl[\frac{1}{1-x}\Bigr]_{+(1)}^{[0,1]} + 1-x \Bigr) + 1-x\right] \theta(0<x<1)
\nonumber \\
&+ 2x\, \mathcal{D}_1(x) - \frac{2x\, \Ein(p|1-x|b)}{|1-x|} - \frac{2|x| \bigl[ \ell_b + E_1(p|x|b)\bigr]}{1-x}
\nonumber \\
&- |1-x|\bigl[\ell_b + E_1(p|1-x|b)\bigr] - (1-x)\, \mathrm{sgn}(x)\bigl[\ell_b + E_1(p|x|b)\bigr] \nonumber \\
&+ \mathrm{sgn}(x)\, e^{-p|x|b} + \frac{e^{-p|x|b} - e^{-p|1-x|b}}{pb(1-x)} \nonumber \\
&\left. - \mathcal{C}_L(x) + \frac{\bigl[1-\cos\bigl(p(1-x)L\bigr)\bigr]\, e^{-p|1-x|b}}{pb(1-x)^2} \right\},
\end{align}
where $P_{qq}(x) = (1+x^2)[1/(1-x)]_{+(1)}^{[0,1]}\theta(0<x<1) + \frac{3}{2}\delta(1-x)$, and
\begin{equation}
\mathcal{C}_L(x) = \cos\bigl(p(1-x)L\bigr)\, \mathcal{D}_0(x).
\end{equation}
The collinear pole and the terms with $\ln(2p/\mu)$ multiply $\mathrm{sgn}(1-x)+\mathrm{sgn}(x)$, which equals 2 for $0<x<1$
and vanishes elsewhere. They therefore have the support of the lightcone PDF, and we write them with the plus distribution on
$[0,1]$, so that the collinear divergence takes the form of the splitting function. The conversion is exact. The
distribution on $[0,2]$ from $x|1-x|^{-1-2\eps}$ combines with the principal value of $x\,\mathrm{sgn}(x)/(1-x)$ from the
second term in the bracket of eq.~\eqref{sailr}, their sum vanishes outside $0<x<1$, and since the principal value over the
interval $[0,2]$, which is symmetric about $x=1$, gives no $\delta(1-x)$, the sum is
$2\,[1/(1-x)]_{+(1)}^{[0,1]}\,\theta(0<x<1)$. The terms in the last four lines of eq.~\eqref{result} extend over the whole
line. Of these, $2x\,\mathcal{D}_1(x)$ and $\mathcal{C}_L(x)$ contain the distributions on $[0,2]$, and in the others the factor
$1/(1-x)$ is the principal value, as in eq.~\eqref{sailr}. The coefficient of $\delta(1-x)$ is
\begin{align} \label{Dcoef}
\mathcal{D} = \frac{\pi^2}{3} &+ 4 + L_b + 2\ln\frac{\mu L}{b_0} + \frac{2L}{b}\arctan\frac{L}{b}
- \ln\Bigl(1+\frac{L^2}{b^2}\Bigr) \nonumber \\
&- \frac{\sin pL}{pL}\ln\Bigl(1+\frac{b^2}{L^2}\Bigr) - \Li_2\Bigl(-\frac{b^2}{L^2}\Bigr)
\nonumber \\
&+ \frac{2}{p}\int_L^{\infty} d\lambda\, \sin(p\lambda) \left[ \frac{1}{\lambda^2+b^2} - \frac{1}{\lambda^2}
-\frac{1}{\lambda^2}\ln\Bigl(1+\frac{b^2}{\lambda^2}\Bigr)\right],
\end{align}
in which $\pi^2/3$ is what remains of the virtual and real sail diagrams at $x=1$, $4 + L_b + 2\ln(\mu L/b_0)$ is
the finite part of the self-energies of the leg and the link, and the remaining terms come from the exchange between the legs
and from $\tilde{B}^{(1)}_{\mathrm{sail,v2}}$. Equation~\eqref{result} is exact in $p^z b_T$ and in $b_T/L$. Its pole part
consists of the collinear divergence $-P_{qq}(x)/\eps$ and a $\delta(1-x)$ pole which depends neither on $p^z$ nor on $L$.
Labeling the infrared part as $-P_{qq}(x)/\eir$ with the help of the scaleless wave-function renormalization, the remaining
pole is $\frac{7}{2}\delta(1-x)/\euv$.

\subsection{Large $L$ and large $p^z b_T$} \label{largeL}

For $L\gg b_T$, the terms in eq.~\eqref{Dcoef} containing $\sin pL$, $\Li_2$ and the integral over $\lambda$ vanish, and the
two distributions in the last line of eq.~\eqref{result} approach
\begin{equation}
-\mathcal{C}_L(x) + \frac{\bigl[1-\cos\bigl(p(1-x)L\bigr)\bigr]\, e^{-p|1-x|b}}{pb(1-x)^2} \longrightarrow
\Bigl[ 2\ln\frac{2pL}{b_0} + \frac{\pi L}{b} - 2 - 2\ln\frac{L}{b}\Bigr] \delta(1-x) + \mathcal{W}(x),
\end{equation}
where $\mathcal{W}(x) = [e^{-p|1-x|b}/(pb(1-x)^2)]_{+(1)}^{\mathbb{R}}$, with the part odd in $1-x$ understood as
a principal value. The logarithms of $L$ come from the self-energies of the legs and from the exchange between them, and
they cancel in pairs,
\begin{equation}
2\ln\frac{2pL}{b_0} - 2\ln\frac{L}{b_T} = 2\ell_b, \qquad 2\ln\frac{\mu L}{b_0} - 2\ln\frac{L}{b_T} = L_b.
\end{equation}
In the first pair the self-energy of the leg of length $L-b^z$ meets the real contribution of the exchange, and in the second
the self-energy of the leg of length $L$ meets the virtual contribution. Each contribution of the exchange also carries
$\pi L/b_T$, and the dependence on $L$ that survives is their sum $2\pi L/b_T$, which is removed by $\sqrt{Z_E(2L,b_T,\mu)}$
in eq.~\eqref{ftilde}.

For $p^z b_T\gg 1$, the terms with $E_1$, $e^{-p|x|b}$ and $\mathcal{W}$ are exponentially suppressed away from $x=0$,
and the remaining dependence on $p^z b_T$ is reorganized by
\begin{equation}
\frac{\Ein(p|1-x|b)}{|1-x|} \longrightarrow \ell_b\, \mathcal{D}_0(x) + \mathcal{D}_1(x) + \Bigl( \ell_b^2 + \frac{\pi^2}{6}
\Bigr) \delta(1-x),
\end{equation}
which follows from $\int_0^T dt\, \Ein(t)/t = \frac{1}{2}\ln^2 T + \gamma_E \ln T + \frac{1}{2}\gamma_E^2 + \pi^2/12 +
\mathcal{O}(e^{-T}/T^2)$. The collinear logarithm $\ell_b$ then combines with $\ln(2p/\mu)$ into $L_b$, and the result becomes
\begin{align} \label{ESZform}
\tilde{B}^{(1)} = \asc &\left\{ -\Bigl(\frac{1}{\eps} + L_b\Bigr) P_{qq}(x) + (1-x)\,\theta(0<x<1) \right. \nonumber \\
&\left. + \delta(1-x) \left[\frac{7}{2\eps} + \frac{9}{2}L_b - \frac{1}{2}L_b^2 - L_bL_z - \frac{1}{2}L_z^2 + L_z
+ \frac{2\pi L}{b_T}\right] \right\},
\end{align}
up to corrections of order $b_T/L$ and $1/(p^zb_T)$. This agrees with eq.~(4.2) of ref.~\cite{Ebert:2019okf}. In
ref.~\cite{Ebert:2019okf} the plus prescription on $[0,1]$ acts on the whole bracket including the term $1-x$, and since
$\int_0^1 dx\,(1-x) = 1/2$, the constant $1/2$ in their $\delta(1-x)$ term is accounted for. The comparison also tests the
contribution of the transverse link in eq.~\eqref{leglink}. It supplies $1/\eps+L_b$ to the coefficient of $\delta(1-x)$, and
without it the coefficients of the pole and of $L_b$ would be $5/2$ and $7/2$ instead of $7/2$ and $9/2$.

Equation~\eqref{ESZform} contains the infrared pole of the lightcone TMDPDF and the dependence on $L$ that is removed by
$\sqrt{Z_E}$, but in the sum of the diagrams the contributions that build the soft factor can no longer be distinguished. The
double logarithms in $L_b$ and $L_z$, for instance, arise partly from the virtual sail, which depends on $p^z$ alone, and
partly from the real sail, which depends on $p^zb_T$. To see which part of the quasi beam function supplies the soft factor,
the integrands have to be separated before the integration over $k_3$ and $\vec{k}_T$.

\section{Regions of the quasi beam function} \label{regions}

We now reorganize the calculation by the method of regions~\cite{Beneke:1997zp}, expanding the integrand of each diagram
before the integration. The leading term of eq.~\eqref{result} in $1/(p^zb_T)$ and $b_T/L$, given in eq.~\eqref{ESZform},
is then reproduced as a sum of contributions from definite momentum regions, and the soft contributions can be compared
directly with the soft limit of the collinear ones.

\subsection{Scales and momentum regions}

The quasi beam function depends on three scales, $p^z$, $1/b_T$ and $1/L$, and on $x$, which is conjugate to $b^z$. We
consider the hierarchy
\begin{equation}
p^z \gg \frac{1}{b_T} \gg \frac{1}{L}, \qquad \lambda = \frac{1}{p^z b_T}\ll 1,
\end{equation}
in which the limit $L\to\infty$ is taken first. With the lightlike vectors $n$ and $\nb$ of section~\ref{intro},
$k_+ = \nb\cdot k$, $k_- = n\cdot k$, and
\begin{equation}
k^{\mu} = \frac{k_+}{2}n^{\mu} + \frac{k_-}{2}\nb^{\mu} + k_T^{\mu}, \qquad p^{\mu} = \frac{p_+}{2}n^{\mu}, \qquad
p_+ = 2p.
\end{equation}
The direction of the longitudinal legs is $\rb^{\mu} = (\nb^{\mu}-n^{\mu})/2$, and accordingly $k_3 = (k_+-k_-)/2$. The momentum
regions that contribute at leading power are
\begin{align}
\text{hard} &: \quad (k_+, k_-, k_T) \sim p\,(1, 1, 1), \nonumber \\
\text{collinear} &: \quad (k_+, k_-, k_T) \sim p\,(1, \lambda^2, \lambda), \nonumber \\
\text{soft} &: \quad (k_+, k_-, k_T) \sim p\,(\lambda, \lambda, \lambda).
\end{align}
The collinear and the soft momenta have the same virtuality $k^2\sim 1/b_T^2$ and differ only in rapidity, which is the
situation of SCET$_{\mathrm{II}}$~\cite{Bauer:2002aj}. This has an important consequence for the regularization of the
individual regions, to which we return in section~\ref{rapreg}.

The phases in the Feynman rules decide which terms survive in each region. In all three regions $|k_3|L\gg 1$, so that the
terms multiplied by $e^{ik_3L}$ oscillate rapidly and vanish as distributions, unless they multiply a singularity at $k_3=0$.
Such singular terms appear only in the Wilson-line diagram. The phase $e^{i\vec{k}_T\cdot\vec{b}_T}$ oscillates in the hard
region and is of order one in the collinear and soft regions. In the real contributions the phase $e^{ik_3b^z}$ fixes
$k_3 = p(1-x)$ after the integration over $b^z$, and the collinear region corresponds to $1-x$ of order one, that is, to $x$ away from $1$, while the soft
region corresponds to $1-x\sim\lambda$. In the soft region $k_3b^z\sim\lambda$
for $b^z\sim 1/p$, and the phase is expanded as $e^{ik_3b^z}\to 1$. The integration over $b^z$ then gives $\delta(1-x)$,
and the soft contributions at leading power are proportional to $\delta(1-x)$, with the integration over $k_3$ unrestricted.

The virtual contribution of the sail diagram has one more region, with $k_3\lesssim 1/L$ and $k_T\sim 1/b_T$, which produces
$\tilde{B}^{(1)}_{\mathrm{sail,v2}}$. It vanishes as $b_T^2/L^2$ or $1/(pL)^2$ and plays no role in the limit $L\to\infty$.

\subsection{Hard region} \label{hardreg}

In the hard region, $k\sim p$, the phases $e^{ik_3L}$ and $e^{i\vec{k}_T\cdot\vec{b}_T}$ in eq.~\eqref{Msail} oscillate
rapidly and can be dropped, and only the first term on the right-hand side of eq.~\eqref{splitid} remains. The integrand
\begin{equation}
ig^2C_F\,\mums\, p\,\delta\bigl(p(1-x)\bigr) \int\frac{d^Dk}{(2\pi)^D}\, \frac{2(p-k_3)-k_-}{(k^2+i0)[(p-k)^2+i0](k_3-i0)}
\end{equation}
is that of $\tilde{B}^{(1)}_{\mathrm{sail,v1}}$, which contains no small scale and needs no further expansion, so that the
hard region of the sail diagram is $\tilde{B}^{(1)}_{\mathrm{sail,v1}}$ in eq.~\eqref{sailh}. The vertex diagram and the real contribution of the sail
diagram contain $e^{i\vec{k}_T\cdot\vec{b}_T}$ and have no hard region at leading power, while the hard region of the
Wilson-line diagram is scaleless, since the diagram does not involve $p$ and its hard region, with the phases dropped,
contains no other scale. The hard contribution at one loop thus comes entirely from the virtual contribution of the sail
diagrams, and in terms of $L_z$ it reads
\begin{equation} \label{hard}
\tilde{B}^{(1)}_{\mathrm{hard}} = \asc\,\delta(1-x) \left[ -\frac{1}{\eps^2} - \frac{1-L_z}{\eps} - \frac{1}{2}L_z^2 + L_z
- 2 + \frac{\pi^2}{12}\right].
\end{equation}
Its finite part is the one-loop matching coefficient $H^{(1)}$ of refs.~\cite{Ebert:2019okf}, \cite[eq.~(25)]{Ji:2019ewn}. The hard region is
defined with its overlaps with the collinear and soft regions removed. These overlaps are scaleless for the same reasons, and when ultraviolet and
infrared poles are distinguished, they cancel the infrared poles of the hard integral, which arise where the gluon becomes soft
or collinear to the quark. The poles in eq.~\eqref{hard} are ultraviolet. They determine the renormalization of $H$, and
together with the ultraviolet poles of the collinear and soft sectors they make up the ultraviolet poles of the quasi beam
function.

\subsection{Collinear region}

For a collinear gluon $k_-\ll k_3\sim k_+$, and the direction of the longitudinal legs is effectively lightlike,
\begin{equation}
\rb\cdot A_n = \frac{1}{2}\bigl(\nb\cdot A_n - n\cdot A_n\bigr) \approx \frac{1}{2}\nb\cdot A_n, \qquad
\frac{1}{k_3-i0} = \frac{2}{k_+-k_- -i0} \approx \frac{2}{k_+-i0}.
\end{equation}
The numerator of the sail diagram is exactly $2(p-k_3)-k_- = p_+-k_+$, and $e^{ik_3b^z}\approx e^{ik_+b^z/2}$.
Dropping the terms with $e^{ik_3L}$, which include the contribution of the transverse link, eq.~\eqref{Msail} becomes
\begin{equation} \label{Msailc}
\tilde{B}^{(1)}_{\mathrm{(a)},c} = 2ig^2C_F\,\mums\, p\int_{-\infty}^{\infty}\frac{db^z}{2\pi}\, e^{-ipb^z(1-x)}
\int\frac{d^Dk}{(2\pi)^D}\, \frac{(p_+-k_+)\bigl(1 - e^{i\vec{k}_T\cdot\vec{b}_T}e^{ik_+b^z/2}\bigr)}
{(k^2+i0)[(p-k)^2+i0](k_+-i0)}.
\end{equation}
The eikonal factor $2/(k_+-i0)$ is the Feynman rule of a Wilson line along $\nb$ built from $\nb\cdot A_n$, and the
integration over $b^z$ gives $2\delta(k_+-p_+(1-x))$, so that $1-x$ is the fraction $k_+/p_+$ carried by the gluon.
Between collinear spinors $\gamma^z$ can be replaced by $\fms{\nb}/2$. Equation~\eqref{Msailc} is the sail diagram of
the lightcone TMD beam function in the Feynman gauge, and the transverse link at $z=L$ decouples from the collinear
gluons in the same way that the transverse link at lightcone infinity decouples in the Feynman gauge. The virtual contribution in
eq.~\eqref{Msailc}, the term without the phase, vanishes. After the integration over $k^0$ its transverse integral is
$\int d^{2-2\eps}k_T/k_T^2$, the integral in eq.~\eqref{Ib} below without the phase $e^{i\vec{k}_T\cdot\vec{b}_T}$. Without
$b_T$ it contains no scale, and in dimensional regularization its ultraviolet and infrared poles cancel.

The real contribution can be evaluated with the $k^0$ integral in eq.~\eqref{k0sail}. For $k_T\ll p$ and $k_3\sim p$, the bracket
multiplying $1/k_T^2$ approaches $\mathrm{sgn}(k_3)+\mathrm{sgn}(p-k_3) = 2\theta(0<x<1)$, with $k_3=p(1-x)$, and the terms with $1/(2E)$
and $1/(2E_q)$ are suppressed by $\lambda^2$. With the transverse integral
\begin{equation} \label{Ib}
4\pi\Bigl(\frac{\mu^2e^{\gamma_E}}{4\pi}\Bigr)^{\eps}\int\frac{d^{2-2\eps}k_T}{(2\pi)^{2-2\eps}}\,
\frac{e^{i\vec{k}_T\cdot\vec{b}_T}}{k_T^2} = \Gamma(-\eps)\Bigl(\frac{\mu^2b_T^2e^{\gamma_E}}{4}\Bigr)^{\eps} \equiv
\mathcal{I}_b = -\frac{1}{\eps} - L_b + \mathcal{O}(\eps),
\end{equation}
the collinear region of the real sail diagrams is
\begin{equation} \label{sailc}
\tilde{B}^{(1)}_{\mathrm{sail},c} = \asc\, \mathcal{I}_b\, \frac{2x}{1-x}\,\theta(0<x<1).
\end{equation}

The other diagram with a collinear region is the vertex diagram, and it is purely collinear. Its hard region is removed by
the phase $e^{i\vec{k}_T\cdot\vec{b}_T}$, and its soft region is suppressed by $k_3^2/p^2$, as can be seen from
eq.~\eqref{Mvertex} with $(p-k)^2\approx -p_+k_-$.
Taking $k_T\ll p$ in the expression in appendix~\ref{app:k0}, we obtain
\begin{equation} \label{vertexc}
\tilde{B}^{(1)}_{\mathrm{vertex},c} = \asc\,(1-\eps)\,\mathcal{I}_b\,(1-x)\,\theta(0<x<1).
\end{equation}
Adding eqs.~\eqref{sailc} and \eqref{vertexc}, the collinear region of the quasi beam function at one loop is
\begin{equation} \label{coll}
\tilde{B}^{(1)}_{c} = \asc\, \mathcal{I}_b \left[ \frac{1+x^2}{1-x} - \eps(1-x)\right] \theta(0<x<1),
\end{equation}
which is the unsubtracted one-loop lightcone TMD beam function. The factor $1/(1-x)$ is not regulated by dimensional
regularization, because the transverse integral in eq.~\eqref{Ib} does not depend on $k_+$.

\subsection{Soft region and zero-bin contribution} \label{softzero}

For a soft gluon the quark propagator becomes eikonal, $(p-k)^2 = k^2-p_+k_-\approx -p_+k_-$, and the numerator
$p_+-k_+$ becomes $p_+$. The quark line acts as a lightlike Wilson line along $n$. The Wilson line is not modified,
since $k_3$, $k_T$ and $1/b_T$ are all of the same order. The attachment to the transverse link is suppressed by $\lambda$
relative to the attachments to the longitudinal legs, because it carries no factor $p/k_3$, and the terms with $e^{ik_3L}$
vanish as before. Since $k_3b^z\sim\lambda$, the phase $e^{ik_3b^z}$ is expanded to $1$, and the integration over $b^z$ gives
$p\,\delta(p(1-x)) = \delta(1-x)$. The soft region of eq.~\eqref{Msail} is
\begin{equation} \label{Msails}
\tilde{B}^{(1)}_{\mathrm{(a)},s} = -ig^2C_F\,\mums\, \delta(1-x)
\int\frac{d^Dk}{(2\pi)^D}\, \frac{1-e^{i\vec{k}_T\cdot\vec{b}_T}}{(k^2+i0)(k_--i0)(k_3-i0)},
\end{equation}
which is the one-gluon exchange between the lightlike line along $n$ and the longitudinal leg of the Wilson line along $\hat{z}$,
with the two quark fields at the same longitudinal position. The term without the phase is scaleless, since the eikonal propagators are homogeneous in $k$ and nothing else sets a
scale. In the soft region the
exchange between the two lightlike lines, which is the soft limit of the vertex diagram, and their self-energies are proportional
to $n^2=0$ and vanish. The Wilson-line diagram involves only the Wilson line and belongs entirely to the soft sector. With
$e^{ik_3b^z}\to 1$ it becomes the one-loop vacuum expectation value of the staple with two legs of equal length $L$. At one
loop the soft sector therefore consists of the exchange between the lightlike line and the Wilson line and of the
self-interaction of the Wilson line. Both are contributions to a vacuum matrix element of the lightlike lines along $n$ and
the staple, which is defined as an operator in section~\ref{operator}.

The soft region is obtained by taking the soft limit of the full integrand. Taking instead the soft limit of the collinear
integrand in eq.~\eqref{Msailc} gives the zero-bin contribution~\cite{Manohar:2006nz}, which is the overlap of the collinear and soft
regions. It is a part of the soft momentum region that is already contained in the collinear region, and it has to be removed
from the collinear region to avoid double counting. Expanding eq.~\eqref{Msailc} for soft momentum, or equivalently expanding
eq.~\eqref{Msails} for $k_-\ll k_3$, gives
\begin{equation} \label{Msail0}
\tilde{B}^{(1)}_{\mathrm{(a)},0} = -2ig^2C_F\,\mums\, \delta(1-x)
\int\frac{d^Dk}{(2\pi)^D}\, \frac{1-e^{i\vec{k}_T\cdot\vec{b}_T}}{(k^2+i0)(k_--i0)(k_+-i0)}.
\end{equation}
The soft region and the zero-bin contribution differ only in the eikonal factor of the leg,
\begin{equation} \label{eikcomp}
\frac{1}{k_3-i0} = \frac{2}{k_+-k_--i0}\quad\text{(soft region)}, \qquad \frac{2}{k_+-i0}\quad\text{(zero-bin contribution)}.
\end{equation}
In the soft region the gluon sees the leg as a line along $\hat{z}$, and in the zero-bin contribution as a line along $\nb$. The two
factors agree in the collinear region, where $k_+\gg k_-$, but not for soft momenta, where $k_+\sim k_-$, and this is why the
zero-bin differs from the soft region. The zero-bin contribution is the one-gluon exchange between lightlike lines
along $n$ and $\nb$, which has the same integrand as the lightcone soft function and as the zero-bin contribution of the lightcone beam
function. In the lightcone case the zero-bin subtraction and the division by the soft function are known to be
equivalent~\cite{Idilbi:2007ff,Idilbi:2007yi}. In the quasi case the zero-bin contribution still has the integrand of the lightcone soft
function, but the soft region does not, since its eikonal factor for the leg is $1/(k_3-i0)$. The two coincide only after the
integration over $k_3$, through the folding of the two hemispheres in section~\ref{foldsec}, where the soft region
reproduces one hemisphere of the lightcone soft function. Since the collinear integrands of the quasi and the lightcone beam
functions coincide, so do their zero-bin contributions, and the purely collinear contributions obtained after the subtraction are
identical.

After the integration over $k^0$, the three regions of the real sail diagrams are obtained from the function
$h(k_3,k_T)$ in eq.~\eqref{k0sail}. With $E=(k_3^2+k_T^2)^{1/2}$, we have
\begin{align} \label{Sregions}
h_c &= \frac{2(p-k_3)}{k_T^2}\,\theta(0<k_3<p), \nonumber \\
h_s &= \frac{p}{k_T^2}\Bigl(1+\frac{k_3}{E}\Bigr) = \frac{p\,k_+}{k_T^2E}, \nonumber \\
h_0 &= \frac{2p}{k_T^2}\,\theta(k_3),
\end{align}
for the collinear region, the soft region and the zero-bin contribution, respectively, where $k_+ = E+k_3$ is evaluated at the pole
$k^0=E$. The zero-bin contribution follows either from $h_c$ for $k_3\ll p$ or from $h_s$ for $k_T\ll|k_3|$, where $E\to|k_3|$ and $1+k_3/E$ becomes
$2\theta(k_3)$, which is eq.~\eqref{eikcomp} after the integration over $k^0$. The soft integrand $h_s/k_3$ is shown in
fig.~\ref{figfold}(a), which is discussed with the folding in section~\ref{foldsec}. The sum $h_c+h_s-h_0$ reproduces $h$ at leading power, up to corrections suppressed by $\lambda$ for soft
$(k_3,k_T)$ and by $\lambda^2$ for collinear $(k_3,k_T)$. The zero-bin contribution is
supported only on $k_3>0$, while the soft region extends over both signs of $k_3$. In the soft region the sail diagram
and its complex conjugate carry $1/(k_3-i0)$ and $1/(k_3+i0)$, and the combination is the principal value in $k_3$. In the
zero-bin contribution the momentum $k_+$ is fixed, and the remaining integration is over $k_-$. For $k_+<0$ the poles of $1/(k^2+i0)$
and $1/(k_--i0)$ lie on the same side of the real axis, and the integral vanishes.

At leading power the one-loop quasi beam function is the sum
\begin{equation} \label{regionsum}
\tilde{B}^{(1)} = \tilde{B}^{(1)}_{\mathrm{hard}} + \bigl(\tilde{B}^{(1)}_c - \tilde{B}^{(1)}_0\bigr) +
\tilde{B}^{(1)}_s,
\end{equation}
in which $\tilde{B}^{(1)}_c - \tilde{B}^{(1)}_0$ is the purely collinear contribution and $\tilde{B}^{(1)}_s$ contains
the soft region of the real sail diagrams and the Wilson-line diagram. The soft region and the zero-bin contribution are proportional to
$\delta(1-x)$, as a consequence of the multipole expansion of $e^{ik_3b^z}$. The assignment of the diagrams to the regions
is summarized in table~\ref{tabregions}.

\begin{table}[t]
\centering
\begin{tabular}{lcccc}
\hline
Diagram & hard & collinear & soft & zero-bin contribution \\
\hline
Sail, virtual & $\bigcirc$ & & & \\
Sail, real & & $\bigcirc$ & $\bigcirc$ & $\bigcirc$ \\
Vertex & & $\bigcirc$ & & \\
Wilson line & & & $\bigcirc$ & \\
Wave function & & & & \\
\hline
\end{tabular}
\caption{\label{tabregions} Regions contributing to each one-loop diagram at leading power in $1/(p^zb_T)$ and $b_T/L$.
The wave-function renormalization is scaleless.}
\end{table}

\subsection{Rapidity divergences} \label{rapreg}

The full result in eq.~\eqref{result} is finite apart from the poles in $\eps$. No additional regulator was needed, since the
eikonal propagators of the Wilson line are spacelike and the quark carries the finite momentum $p^z$. The individual regions,
however, are not finite in the same sense. The collinear region in eq.~\eqref{coll} behaves as $1/(1-x)$ at $x\to 1$, where $k_+\to 0$,
for any $\eps$. The soft region and the zero-bin contribution are proportional to $\delta(1-x)$, and the integration over $k_3$ that
multiplies it diverges logarithmically. For the soft region the divergence comes from $k_3\to\infty$, where
$h_s/k_3\to 2p/(k_T^2k_3)$, while for $k_3\to-\infty$ the integrand falls off as $1/k_3^3$. For the
zero-bin contribution, $\int_0^{\infty}dk_3/k_3$ diverges at both ends. These are rapidity divergences, characteristic of collinear
and soft modes with the same virtuality, and they require a regulator beyond dimensional regularization.

The roles of the two regulators are most transparent if the integrals are organized in a definite order, although the result
does not depend on it. The integration over $k^0$, equivalently over $k_-$ at fixed $k_3$, is performed first by contour
integration and produces no divergence. The transverse integration at fixed $k_3$ is performed next, and all poles in $\eps$
arise there. In the real contributions they appear through $\mathcal{I}_b$ in eq.~\eqref{Ib}, multiplied by a function of
$k_3$. The remaining integration over $k_3$, or over $k_+$, produces the rapidity divergences, which are regulated by
the additional regulator. In the collinear sector they multiply $\mathcal{I}_b$, and the poles in $\eps$ and $\eta$ stay
separate. In the soft sector the hemisphere regulator introduced below makes the integral over $k_+$ depend on $k_T$, and the
factor $k_T^{-\eta}$ enters the transverse integral. The limit $\eta\to 0$ is then taken before $\eps\to 0$, and the
rapidity pole combined with the transverse integral produces a double pole $1/\eps^2$, which cancels the double pole of the
hard region in the sum of the regions.

The rapidity divergences can be labeled in the same way as the poles in $\eps$. With a regulator of the form
$(\nu/k_+)^{\eta}$, a pole from large $k_+$ is ultraviolet and one from small $k_+$ is infrared, and we denote them by
$1/\eta_{\mathrm{UV}}$ and $1/\eta_{\mathrm{IR}}$. As in dimensional regularization, a scaleless integral vanishes through the
cancellation of the two,
\begin{equation} \label{etauvir}
\int_0^{\infty}\frac{dk_+}{k_+}\Bigl(\frac{\nu}{k_+}\Bigr)^{\eta} = \int_0^{\nu}\frac{dk_+}{k_+}\Bigl(\frac{\nu}{k_+}
\Bigr)^{\eta} + \int_{\nu}^{\infty}\frac{dk_+}{k_+}\Bigl(\frac{\nu}{k_+}\Bigr)^{\eta} = -\frac{1}{\eta_{\mathrm{IR}}} +
\frac{1}{\eta_{\mathrm{UV}}},
\end{equation}
where the first integral converges for $\eta<0$ and the second for $\eta>0$.

It is often stated that rapidity divergences arise when $k_+$ approaches either zero or infinity. This is not the case for
a sector defined with its overlap removed. In the collinear region the gluon momentum is bounded, $0<k_+<p_+$, and a pole
arises only from $k_+\to 0$, where the gluon has left the collinear region and entered the soft one. The zero-bin contribution, whose
integrand coincides with the collinear one at small $k_+$, contains the same infrared pole, and its subtraction leaves the
ultraviolet pole of the zero-bin contribution from large $k_+$. In the soft region the hemisphere $k_3>0$ is bounded from below by
$k_+>k_T$, and the integrand falls off as $1/k_3^3$ in the hemisphere $k_3<0$, so that the only rapidity divergence comes
from $k_+\to\infty$, where the soft gluon approaches the collinear direction. Each sector thus contains only an ultraviolet
rapidity divergence, and the two poles cancel in the sum of the sectors, since the quasi beam function has no rapidity
divergence. The origin of each pole is summarized in table~\ref{tabrap}, with the coefficients obtained in
section~\ref{sectors}.

\begin{table}[t]
\centering
\setlength{\tabcolsep}{4pt}
\begin{tabular}{l>{\raggedright\arraybackslash}p{2.2cm}>{\raggedright\arraybackslash}p{3.5cm}l}
\hline
Contribution & Range of $k_+$ & Origin of the pole & Pole \\
\hline
Collinear region & $0<k_+<p_+$ & $k_+\to 0$ & $-1/\eta_{\mathrm{IR}}$ \\
Zero-bin contribution & $0<k_+<\infty$ & $k_+\to 0$ and $k_+\to\infty$ & $1/\eta_{\mathrm{UV}}-1/\eta_{\mathrm{IR}}$ \\
Collinear sector & & $k_+\to\infty$, from the zero-bin contribution & $-1/\eta_{\mathrm{UV}}$ \\
Soft sector & both hemispheres & $k_+\to\infty$ for $k_3>0$ & $+1/\eta_{\mathrm{UV}}$ \\
\hline
\end{tabular}
\caption{\label{tabrap} Rapidity poles of the regions and sectors, in units of $2\,\mathcal{I}_b\,\delta(1-x)\,\as C_F/(2\pi)$.
The collinear sector is the collinear region after the zero-bin subtraction. Only ultraviolet poles remain in the sectors, and
they cancel in the sum.}
\end{table}

We use the rapidity regulator in the hemisphere form of ref.~\cite{Chay:2020jzn},
\begin{equation} \label{RS}
R_S = \Bigl(\frac{\nu_+}{k_+}\Bigr)^{\eta_+}\Theta(k_+-k_-) + \Bigl(\frac{\nu_-}{k_-}\Bigr)^{\eta_-}\Theta(k_--k_+),
\end{equation}
which is a version of the $\eta$ regulator of ref.~\cite{Chiu:2012ir} applied to the gluon momentum. Since
$k_+-k_- = 2k_3$, the hemisphere is $\Theta(k_+-k_-) = \Theta(k_3)$, and the boundary between the two hemispheres is the
plane $k_3=0$, perpendicular to the longitudinal legs. For $n$-collinear momenta $k_+\gg k_-$, only the first term contributes, and with
$k_+ = 2k_3 = p_+(1-x)$ the regulator multiplies the collinear region by
\begin{equation}
\Bigl(\frac{\nu_+}{p_+}\Bigr)^{\eta_+}(1-x)^{-\eta_+}.
\end{equation}
In the collinear sector the regulator in terms of $k_3$ and that in terms of $k_+$ coincide at leading power.

In the soft region the regulator acts in both hemispheres. The eikonal propagator of the spacelike leg is singular at $k_3=0$,
which is the boundary between the hemispheres, as seen in fig.~\ref{figfold}(a). The principal value pairs $k_3$ with
$-k_3$ across this boundary, and it is well defined only if $R_S$ takes the same value on both sides. At $k_3=0$ the gluon has $k_+=k_-=k_T$, and continuity requires $\eta_+=\eta_-$ and $\nu_+=\nu_-$.
With different parameters the discontinuity of $R_S$ would produce a logarithmic divergence at $k_3\to 0$. The
symmetric choice reflects the fact that the eikonal propagator of the longitudinal legs is singular on the plane that separates
the two hemispheres, and we use
$\eta_\pm=\eta$ and $\nu_\pm=\nu$ in the following. The same regulator is placed on the lightcone soft function, with which
the soft region is compared in section~\ref{sectors}. There the hemisphere boundary $k_+>k_-$ with $k_+k_-=k_T^2$
becomes $k_+>k_T$, which is the origin of the factor $k_T^{-\eta}$ discussed above.

In the multipole expansion the zero-bin contribution is
\begin{equation} \label{zerobin}
\tilde{B}^{(1)}_0 = 2\asc\,\mathcal{I}_b\,\delta(1-x)\int_0^{\infty}\frac{dk_3}{k_3}\Bigl(\frac{\nu}{2k_3}
\Bigr)^{\eta} = 2\asc\,\mathcal{I}_b\,\delta(1-x)\Bigl(\frac{1}{\eta_{\mathrm{UV}}}-\frac{1}{\eta_{\mathrm{IR}}}\Bigr),
\end{equation}
and the explicit results for the two sectors are presented in the next section.

\section{Collinear and soft sectors at one loop} \label{sectors}

We now evaluate the regions of section~\ref{regions} with the rapidity regulator in eq.~\eqref{RS}, using the symmetric
choice $\eta_\pm=\eta$ and $\nu_\pm=\nu$. The hard region was already given in eq.~\eqref{hard}, and it needs no rapidity
regulator.

\subsection{Collinear sector}

With the regulator, the collinear region of the real sail diagrams in eq.~\eqref{sailc} and that of the vertex diagram in
eq.~\eqref{vertexc} become
\begin{equation}
\tilde{B}^{(1)}_c = \asc\,\mathcal{I}_b\Bigl(\frac{\nu}{p_+}\Bigr)^{\eta}\Bigl[ 2x(1-x)^{-1-\eta} + (1-\eps)(1-x)^{1-\eta}
\Bigr]\theta(0<x<1).
\end{equation}
The factor $(1-x)^{-1-\eta}$ is expanded as a distribution,
\begin{equation}
(1-x)^{-1-\eta}\,\theta(0<x<1) = -\frac{1}{\eta}\,\delta(1-x) + \Bigl[\frac{1}{1-x}\Bigr]_{+(1)}^{[0,1]}\theta(0<x<1) +
\mathcal{O}(\eta),
\end{equation}
where the pole comes from $x\to 1$, that is, from $k_+\to 0$, and is therefore an infrared rapidity divergence. With
$2x[1/(1-x)]_{+(1)}^{[0,1]} + 1-x = (1+x^2)[1/(1-x)]_{+(1)}^{[0,1]}$, the collinear region is
\begin{align}
\tilde{B}^{(1)}_c = \asc\,\mathcal{I}_b &\left\{ -2\,\delta(1-x)\Bigl[\frac{1}{\eta_{\mathrm{IR}}} + \ln\frac{\nu}{p_+}\Bigr]
\right. \nonumber \\
&\left. + \Bigl[(1+x^2)\Bigl[\frac{1}{1-x}\Bigr]_{+(1)}^{[0,1]} - \eps(1-x)\Bigr]\theta(0<x<1)\right\}.
\end{align}
Subtracting the zero-bin contribution in eq.~\eqref{zerobin} replaces $1/\eta_{\mathrm{IR}}$ by $1/\eta_{\mathrm{UV}}$, and the purely
collinear contribution is
\begin{align} \label{Bn}
B_n^{(1)}(x) \equiv \tilde{B}^{(1)}_c - \tilde{B}^{(1)}_0 = \asc &\left\{ -\Bigl(\frac{1}{\eps}+L_b\Bigr)\Bigl[P_{qq}(x) -
\frac{3}{2}\delta(1-x)\Bigr] + (1-x)\,\theta(0<x<1) \right. \nonumber \\
&\left. - 2\,\mathcal{I}_b\Bigl[\frac{1}{\eta_{\mathrm{UV}}} + \ln\frac{\nu}{p_+}\Bigr]\delta(1-x)\right\},
\end{align}
where we have used $\mathcal{I}_b = -1/\eps-L_b+\mathcal{O}(\eps)$ and $-\eps\,\mathcal{I}_b = 1+\mathcal{O}(\eps)$. In
the last term $\mathcal{I}_b$ is kept unexpanded, since $\eta\to 0$ is taken before $\eps\to 0$ and the coefficient of
$1/\eta$ is needed to all orders in $\eps$. Since
the collinear integrands of the quasi and the lightcone beam functions coincide, eq.~\eqref{Bn} is the one-loop TMDPDF
beam function with the regulator $R_S$, and it depends on $p^z$ only through $\ln(\nu/p_+)$.

\subsection{Soft region of the real sail diagrams} \label{softsec}

In the multipole expansion the soft region of the real sail diagrams is obtained from $h_s$ in
eq.~\eqref{Sregions}. Replacing a function of $p(1-x)$ by $\delta(1-x)$ times its integral over $x$, we have
\begin{equation} \label{softk3}
\tilde{B}^{(1)}_{\mathrm{sail},s} = \asc\,\delta(1-x)\int[dk_T]\,\frac{e^{i\vec{k}_T\cdot\vec{b}_T}}{k_T^2}
\int_{-\infty}^{\infty}dk_3\,\PV\frac{1}{k_3}\,\frac{k_+}{E}\,R_S,
\end{equation}
with the transverse measure
\begin{equation} \label{dkT}
\int[dk_T]\equiv 4\pi\Bigl(\frac{\mu^2e^{\gamma_E}}{4\pi}\Bigr)^{\eps}\int\frac{d^{2-2\eps}k_T}{(2\pi)^{2-2\eps}},
\end{equation}
which is that of eq.~\eqref{Ib}, so that $\int[dk_T]\,e^{i\vec{k}_T\cdot\vec{b}_T}/k_T^2=\mathcal{I}_b$. The gluon is on shell at the pole $k^0=E$, so that $k_\pm = E\pm k_3$
and $k_+k_- = k_T^2$. At fixed $k_T$ we change the variable from $k_3$ to $k_+$, using
\begin{equation}
k_3 = \frac{k_+^2-k_T^2}{2k_+}, \qquad dk_3 = \frac{E}{k_+}\,dk_+.
\end{equation}
The range $-\infty<k_3<\infty$ is mapped onto $0<k_+<\infty$, and the hemisphere boundary $k_3=0$ onto $k_+=k_T$.
The integral over $k_3$ in eq.~\eqref{softk3} becomes
\begin{equation} \label{softkp}
\PV\int_0^{\infty}dk_+\,\frac{2k_+}{k_+^2-k_T^2}\,R_S = \int_{k_T}^{\infty}dk_+\,\frac{2k_+}{k_+^2-k_T^2}
\Bigl(\frac{\nu}{k_+}\Bigr)^{\eta} + \PV\int_0^{k_T}dk_+\,\frac{2k_+}{k_+^2-k_T^2}\Bigl(\frac{\nu}{k_-}\Bigr)^{\eta}.
\end{equation}
\subsection{Folding of the hemispheres} \label{foldsec}

\begin{figure}[t]
\begin{center}
\includegraphics[width=\textwidth]{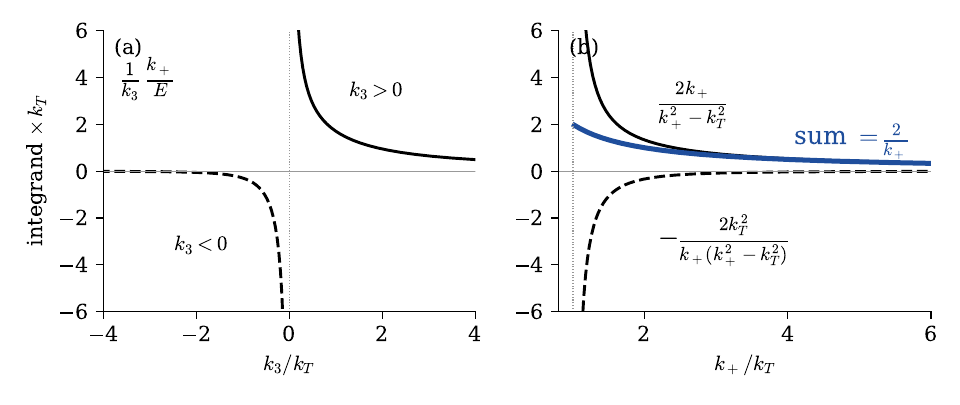}
\end{center}
\caption{\label{figfold} Integrand of the soft region of the real sail diagrams at fixed $k_T$, shown without the rapidity regulator. (a) As a function of $k_3$,
for the hemispheres $k_3>0$ (solid) and $k_3<0$ (dashed), whose contributions are paired by the principal value. (b) After the
change of variable to $k_+$ and the substitution $k_+\to k_T^2/k_+$ in the hemisphere $k_3<0$, both hemispheres lie on
$k_+>k_T$. Their singularities at $k_+=k_T$ cancel, and the sum (thick blue line) is the integrand $2/k_+$ of the lightcone
soft function in one hemisphere.}
\end{figure}

The first integral in eq.~\eqref{softkp} covers the hemisphere $k_3>0$, where $k_+>k_T$, and the second covers $k_3<0$,
where $k_+<k_T$. They share the factor $2k_+/(k_+^2-k_T^2)$, but neither has the integrand of the lightcone soft function,
$2/k_+$, and they differ in the regulator, which depends on $k_+$ in the first hemisphere and on $k_-$ in the second.
Reflecting $k_3\to -k_3$ leaves $E$ and $k_T$ unchanged and interchanges $k_+$ and $k_-$. In the second integral we therefore
use $k_-$ as the variable, writing $q=k_-=k_T^2/k_+$, so that
\begin{equation}
k_+ = \frac{k_T^2}{q}, \qquad dk_+ = -\frac{k_T^2}{q^2}\,dq,
\end{equation}
and the range $0<k_+<k_T$ becomes $k_T<q<\infty$. The regulator $(\nu/k_-)^{\eta}$ becomes $(\nu/q)^{\eta}$, and the second
integral becomes
\begin{equation}
-\int_{k_T}^{\infty}dq\,\frac{2k_T^2}{q(q^2-k_T^2)}\Bigl(\frac{\nu}{q}\Bigr)^{\eta}.
\end{equation}
The variable $q$ is a dummy variable, and calling it $k_+$ again, both integrals run over $k_T<k_+<\infty$ with the same
regulator $(\nu/k_+)^{\eta}$. Their integrands can then be added, and since
\begin{equation}
\frac{2k_+}{k_+^2-k_T^2} - \frac{2k_T^2}{k_+(k_+^2-k_T^2)} = \frac{2(k_+^2-k_T^2)}{k_+(k_+^2-k_T^2)} = \frac{2}{k_+},
\end{equation}
the two hemispheres combine into
\begin{equation} \label{fold}
\PV\int_0^{\infty}dk_+\,\frac{2k_+}{k_+^2-k_T^2}\,R_S = \int_{k_T}^{\infty}dk_+\,\frac{2}{k_+}\Bigl(\frac{\nu}{k_+}
\Bigr)^{\eta}.
\end{equation}
We refer to this mapping of the hemisphere $k_3<0$ onto $k_3>0$, followed by the sum of the two, as the folding of the
hemispheres. Only the sum of the two hemispheres has the integrand of the lightcone soft function, as shown in
fig.~\ref{figfold}. The singularities at $k_+=k_T$ of the two terms cancel, and eq.~\eqref{fold} holds for any $\eta$ and at every $k_T$.
The equality requires the symmetric regulator, since the substitution exchanges $(\nu_-/k_-)^{\eta_-}$ with
$(\nu_+/k_+)^{\eta_+}$.

The right-hand side of eq.~\eqref{fold} is the integrand of the lightcone soft function. Its one-loop real contribution,
the exchange between the lines along $n$ and $\nb$, has the integrand of eq.~\eqref{Msail0} together with its complex
conjugate, and is
\begin{equation}
S^{(1)}_{n\nb} = 2\asc\int[dk_T]\,\frac{e^{i\vec{k}_T\cdot\vec{b}_T}}{k_T^2}\int_0^{\infty}\frac{dk_+}{k_+}\,R_S
= S_+ + S_-,
\end{equation}
where the hemisphere regulator $R_S$ splits the integral into
\begin{align}
S_+ &= 2\asc\int[dk_T]\,\frac{e^{i\vec{k}_T\cdot\vec{b}_T}}{k_T^2}\int_{k_T}^{\infty}\frac{dk_+}{k_+}\Bigl(\frac{\nu}{k_+}\Bigr)^{\eta},
\nonumber \\
S_- &= 2\asc\int[dk_T]\,\frac{e^{i\vec{k}_T\cdot\vec{b}_T}}{k_T^2}\int_0^{k_T}\frac{dk_+}{k_+}\Bigl(\frac{\nu}{k_-}\Bigr)^{\eta}.
\end{align}
The substitution $k_+=k_T^2/q$ used in the folding maps $S_-$ onto $S_+$, which expresses the symmetry of $S_{n\nb}$ under the
interchange of $n$ and $\nb$, and with the symmetric regulator $S_-=S_+$. The virtual contributions are scaleless, since
without the phase they consist of eikonal propagators alone. Inserting eq.~\eqref{fold} into eq.~\eqref{softk3}, the $k_+$
integral is that of $S_+$, and
\begin{equation} \label{softSplus}
\tilde{B}^{(1)}_{\mathrm{sail},s} = S_+\,\delta(1-x) = \frac{1}{2}S^{(1)}_{n\nb}\,\delta(1-x).
\end{equation}
Through the folding, the puzzle raised in the Introduction can be explained at one loop, namely how the soft region, which
contains the single lightlike direction $n$, can account for the soft function with the two lightlike directions $n$ and
$\nb$. The reflection $k_3\to -k_3$, which maps one hemisphere of the gluon momentum onto the other, interchanges $k_+$ and
$k_-$, and with them the roles of $n$ and $\nb$, so that the spacelike direction $\hat{z}$ takes the place of $\nb$. Beyond
one loop the equality holds for the rapidity divergences, as shown in section~\ref{operator}, while the remainder enters
$S_r$.

\subsection{Soft sector at one loop} \label{softone}

To evaluate $S_+$ we perform the integration over $k_+$ first,
$\int_{k_T}^{\infty}dk_+\,k_+^{-1-\eta}\nu^{\eta} = (\nu/k_T)^{\eta}/\eta$, and then the transverse integral, which is
\begin{equation}
\int[dk_T]\,\frac{e^{i\vec{k}_T\cdot\vec{b}_T}}{(k_T^2)^{1+\eta/2}} = \Bigl(\frac{b_T}{2}\Bigr)^{\eta}
\Bigl(\frac{\mu^2b_T^2e^{\gamma_E}}{4}\Bigr)^{\eps}\frac{\Gamma(-\eps-\eta/2)}{\Gamma(1+\eta/2)}.
\end{equation}
Since the factor $k_T^{-\eta}$ enters the transverse integral, we expand in $\eta$ before $\eps$, and obtain
\begin{align} \label{Splus}
S_+ &= \frac{2}{\eta}\asc\Bigl(\frac{\nu b_T}{2}\Bigr)^{\eta}\Bigl(\frac{\mu^2b_T^2e^{\gamma_E}}{4}\Bigr)^{\eps}
\frac{\Gamma(-\eps-\eta/2)}{\Gamma(1+\eta/2)} \nonumber \\
&= 2\asc\left[ \frac{\mathcal{I}_b}{\eta_{\mathrm{UV}}} + \mathcal{I}_b\ln\frac{\nu b_T}{b_0} +
\frac{1}{2\eps^2} + \frac{L_b}{2\eps} + \frac{L_b^2}{4} - \frac{\pi^2}{24}\right] + \mathcal{O}(\eta,\eps).
\end{align}
The pole in $\eta$ comes from $k_+\to\infty$ and is ultraviolet in origin.

\begin{figure}[t]
\begin{center}
\includegraphics[width=\textwidth]{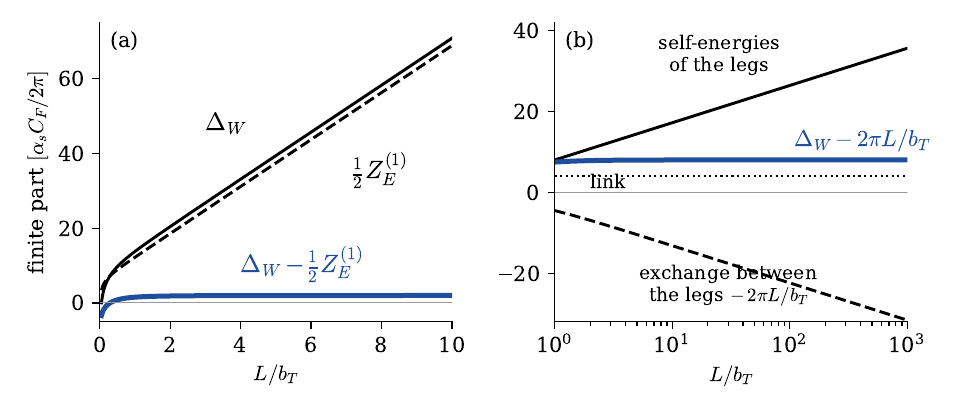}
\end{center}
\caption{\label{figL} Dependence on $L$ of the self-interaction of the Wilson line, with the finite parts in units of
$\as C_F/(2\pi)$ at $\mu b_T = e\,b_0$, where $L_b=2$. (a) $\Delta_W$ in eq.~\eqref{DeltaW} and half of the rectangular
Wilson loop $Z_E^{(1)}$ grow linearly in $L$ with the same slope, and their difference (blue) approaches $L_b$, the
contribution of one straight segment. (b) With the linear term $2\pi L/b_T$ removed, the self-energies of the legs grow as
$4\ln(L/b_T)$ and the exchange between the legs falls as $-4\ln(L/b_T)$, while the transverse link does not depend on $L$.
The sum (blue) approaches $3L_b+2$.}
\end{figure}

The other part of the soft sector is the Wilson-line diagram. With $e^{ik_3b^z}\to 1$ it is the one-loop vacuum expectation
value of the staple with two legs of length $L$, and it needs no rapidity regulator. From the results of
section~\ref{quasi}, the self-energies of the two legs and of the transverse link and the exchange between the legs give
\begin{align}
\Delta_W &= \asc\left\{ 2\Bigl[\frac{1}{\eps} + 2\ln\frac{\mu L}{b_0} + 2\Bigr] +
\Bigl[\frac{1}{\eps} + L_b + 2\Bigr] \right. \nonumber \\
&\qquad\left. + 2\Bigl[\frac{2L}{b_T}\arctan\frac{L}{b_T} - \ln\Bigl(1+\frac{L^2}{b_T^2}\Bigr)\Bigr]\right\} \nonumber \\
&= \asc\left[ \frac{3}{\eps} + 3L_b + 2 + \frac{2\pi L}{b_T}\right] + \mathcal{O}\Bigl(\frac{b_T}{L}\Bigr).
\label{DeltaW}
\end{align}
In the first line the self-energies of the two legs, both of length $L$ once $b^z=0$, and of the transverse link of length
$b_T$ are given by eq.~\eqref{selfen}. The second line is the exchange between the legs, which appears twice because it has a
virtual and a real contribution. They come from the two ends of the leg at $\vec{b}_T$, the end at $z=L$ and the end at
$z=b^z$, and setting $b^z=0$ makes them equal. For $L\gg b_T$ the logarithms of $L$ in the self-energies of the legs cancel
against those in the exchange, as in section~\ref{largeL} and fig.~\ref{figL}(b), and what remains of the dependence on $L$
is the linear term $2\pi L/b_T$. The soft sector at one loop is therefore
\begin{equation} \label{softsector}
\tilde{B}^{(1)}_s = \bigl(S_+ + \Delta_W\bigr)\delta(1-x).
\end{equation}

\subsection{Sum of the regions}

In the sum of the collinear and the soft sectors the ultraviolet rapidity poles cancel. The coefficient of
$1/\eta_{\mathrm{UV}}$ is $-2\mathcal{I}_b$ in eq.~\eqref{Bn} and $2\mathcal{I}_b$ in eq.~\eqref{Splus}. Since both are the same function $\mathcal{I}_b$ of
$\eps$, the rapidity poles cancel to all orders in $\eps$. The dependence on $\nu$ cancels in the same way. The remaining dependence on $p^z$ and $b_T$ combines through
$\ln p_+ + \ln(b_T/b_0) = \ell_b = (L_b+L_z)/2$, and we obtain
\begin{align} \label{BnSplus}
B_n^{(1)} + S_+\delta(1-x) = \asc &\left\{ -\Bigl(\frac{1}{\eps}+L_b\Bigr)\Bigl[P_{qq}(x)-\frac{3}{2}\delta(1-x)\Bigr]
+ (1-x)\,\theta(0<x<1) \right. \nonumber \\
&\left. + \delta(1-x)\Bigl[ \frac{1}{\eps^2} - \frac{L_z}{\eps} - \frac{1}{2}L_b^2 - L_bL_z - \frac{\pi^2}{12}\Bigr]\right\}.
\end{align}
Adding the hard contribution in eq.~\eqref{hard}, the double pole and the pole proportional to $L_z$ cancel, and the
coefficient of $\delta(1-x)$ becomes $-1/\eps - \frac{1}{2}L_b^2 - L_bL_z - \frac{1}{2}L_z^2 + L_z - 2$. Adding $\Delta_W$
in eq.~\eqref{DeltaW} and writing $P_{qq}$ in full, we obtain
\begin{align} \label{regionresult}
\tilde{B}^{(1)} = \asc &\left\{ -\Bigl(\frac{1}{\eps} + L_b\Bigr) P_{qq}(x) + (1-x)\,\theta(0<x<1) \right. \nonumber \\
&\left. + \delta(1-x) \left[\frac{7}{2\eps} + \frac{9}{2}L_b - \frac{1}{2}L_b^2 - L_bL_z - \frac{1}{2}L_z^2 + L_z
+ \frac{2\pi L}{b_T}\right] \right\},
\end{align}
which is eq.~\eqref{ESZform}. The leading-power result of the full calculation is reproduced by
\begin{equation} \label{factB}
\tilde{B}^{(1)} = \tilde{B}^{(1)}_{\mathrm{hard}} + B_n^{(1)} + \bigl(S_+ + \Delta_W\bigr)\delta(1-x),
\end{equation}
in which the hard term comes from the virtual sail diagram, $B_n^{(1)}$ is the lightcone TMD beam function, and
$S_++\Delta_W$ is the soft limit of the quasi beam function, in which the quark supplies a lightlike Wilson line along $n$ and
the Wilson line is kept intact. Each of the three terms is a definite part of the quasi beam function, and in the next section
this decomposition is related to the factorization formula for the quasi-TMDPDF.

\section{Soft factor and the factorization formula} \label{softfactor}

The quasi-TMDPDF in eq.~\eqref{ftilde} is related to the TMDPDF $f^{\mathrm{TMD}}(x,b_T,\mu,\zeta)$ by the factorization
formula of the Introduction~\cite{Ji:2019sxk}, \cite[eq.~(22)]{Ji:2019ewn}
\begin{equation} \tag{\ref{factintro}}
\tilde{f}(x,b_T,\mu,\zeta_z)\sqrt{S_r(b_T,\mu)} = H\Bigl(\frac{\zeta_z}{\mu^2}\Bigr)\,
e^{\frac{1}{2}K(b_T,\mu)\ln(\zeta_z/\zeta)} f^{\mathrm{TMD}}(x,b_T,\mu,\zeta),
\end{equation}
up to corrections suppressed by $1/(P^zb_T)$, where $\zeta_z = (2xP^z)^2$, $K$ is the Collins-Soper kernel, and $S_r$ is the
reduced soft function. It is equivalent to the formulation of ref.~\cite{Ebert:2019okf}, in which the soft factor is
absorbed into the quasi-TMDPDF through a quasi soft function. For the external quark the leading-power region is $x\to 1$, and we set $\zeta_z = (2p^z)^2 = p_+^2$, so
that $L_z = \ln(\zeta_z/\mu^2)$. Once the quasi beam function is divided by $\sqrt{Z_E}$,
the decomposition in eq.~\eqref{factB} reproduces eq.~\eqref{factintro} at one loop, and each factor of the formula can be
traced to a definite part of the quasi beam function.

\subsection{Rectangular Wilson loop}

The factor $Z_E(2L,b_T,\mu)$ in eq.~\eqref{ftilde} is the vacuum expectation value of a rectangular Wilson loop with sides
$2L$ and $b_T$. At one loop it is obtained from the same ingredients as the Wilson-line diagram of the quasi beam function.
The self-energy of each side is given by eq.~\eqref{selfen}, with the length $2L$ for the two long sides and $b_T$ for the
two short sides. The exchange between the two long sides, which are antiparallel, is that between the two legs of the Wilson line
in eq.~\eqref{DeltaW} with $L$ replaced by $2L$. The exchange between the two short sides is of order $b_T^2/L^2$, and the
exchanges between adjacent sides vanish in the Feynman gauge, since the sides are perpendicular. We obtain
\begin{align}
Z_E^{(1)} &= \asc\left\{ 2\Bigl[\frac{1}{\eps} + 2\ln\frac{2\mu L}{b_0} + 2\Bigr] + 2\Bigl[\frac{1}{\eps} + L_b + 2\Bigr]
\right. \nonumber \\
&\qquad\left. + 2\Bigl[\frac{4L}{b_T}\arctan\frac{2L}{b_T} - \ln\Bigl(1+\frac{4L^2}{b_T^2}\Bigr)\Bigr]\right\} \nonumber \\
&= \asc\left[ \frac{4}{\eps} + 4L_b + 4 + \frac{4\pi L}{b_T}\right] + \mathcal{O}\Bigl(\frac{b_T}{L}\Bigr).
\end{align}
As in the Wilson-line diagram, the logarithms of $L$ from the self-energies of the long sides cancel against those from the exchange.
Comparing with eq.~\eqref{DeltaW}, we find
\begin{equation} \label{DWZE}
\Delta_W - \frac{1}{2}Z_E^{(1)} = \asc\Bigl(\frac{1}{\eps} + L_b\Bigr).
\end{equation}
Half of the loop enters through $\sqrt{Z_E}$, and at one loop it consists of one long side of length $2L$ and one short side
of length $b_T$, together with half of the exchange between the long sides. The transverse link and the short side have the
same length $b_T$ and cancel in the difference, which therefore compares the two legs of length $L$ with the long side of
length $2L$. With $\Sigma(\ell)$ in eq.~\eqref{selfen} and
$X(\ell)=(2\ell/b_T)\arctan(\ell/b_T)-\ln(1+\ell^2/b_T^2)$ the exchange between antiparallel lines of length $\ell$,
\begin{align}
\Delta_W - \frac{1}{2}Z_E^{(1)} &= \asc\Bigl\{\bigl[2\Sigma(L)-\Sigma(2L)\bigr] + \bigl[2X(L)-X(2L)\bigr]\Bigr\} \nonumber \\
&= \asc\Bigl\{\Bigl[\frac{1}{\eps}+2\ln\frac{\mu L}{b_0}+2-2\ln 2\Bigr]
-\Bigl[2+2\ln\frac{L}{b_T}-2\ln 2\Bigr]\Bigr\},
\end{align}
up to corrections of order $b_T/L$. Cutting a straight line in two adds the ultraviolet divergence of one more segment, and
the exchange removes the logarithms of $L$ and the constants, so that the difference is the contribution of one straight
segment.

The cancellation of the term linear in $L$, shown in fig.~\ref{figL}(a), is the one-loop form of the cancellation of the static
potential discussed in section~\ref{quasi}. It takes place because half of the loop with sides $2L$ and $b_T$ has antiparallel sides of the same length as the
staple.

\subsection{Matching and evolution}

Dividing eq.~\eqref{factB} by $\sqrt{Z_E}$, the quasi-TMDPDF at one loop is
\begin{equation} \label{ftildedec}
\tilde{f}^{(1)} = \tilde{B}^{(1)}_{\mathrm{hard}} + \bigl[B_n^{(1)} + S_+\,\delta(1-x)\bigr] + \Bigl(\Delta_W -
\frac{1}{2}Z_E^{(1)}\Bigr)\delta(1-x).
\end{equation}
The bracket is the TMDPDF. In the definition $f^{\mathrm{TMD}} = B_n\sqrt{S_{n\nb}}$ the one-loop contribution of
the soft function is $\frac{1}{2}S^{(1)}_{n\nb} = S_+$, and the bracket is given by eq.~\eqref{BnSplus}. Its dependence on
$\nu$ has cancelled, and its dependence on $p^z$ enters through $\ln p_+$ in eq.~\eqref{Bn}, which identifies the rapidity
scale as $\zeta = p_+^2 = \zeta_z$. After $\msbar$ renormalization the coefficient of $\delta(1-x)$ in $f^{\mathrm{TMD}}$ is
\begin{equation}
\asc\Bigl[ \frac{3}{2}L_b - \frac{1}{2}L_b^2 - L_bL_z - \frac{\pi^2}{12}\Bigr],
\end{equation}
which agrees with eq.~(4.1) of ref.~\cite{Ebert:2019okf} at $\zeta=\zeta_z$, once their plus prescription, which acts on the
term $1-x$ as well, is taken into account.

With the rapidity scale $\zeta=\zeta_z$ identified above, the exponential factor in eq.~\eqref{factintro} equals one, and at
one loop the formula reads
\begin{equation}
\tilde{f}^{(1)} = H^{(1)}\delta(1-x) + f^{\mathrm{TMD}(1)} - \frac{1}{2}S_r^{(1)}\delta(1-x).
\end{equation}
Comparing with eq.~\eqref{ftildedec}, the finite part of the hard contribution in eq.~\eqref{hard} gives
\begin{equation}
H^{(1)} = \asc\Bigl( -\frac{1}{2}L_z^2 + L_z - 2 + \frac{\pi^2}{12}\Bigr),
\end{equation}
and eq.~\eqref{DWZE} gives, after $\msbar$ renormalization,
\begin{equation} \label{Sr1}
S_r^{(1)} = -2\Bigl(\Delta_W - \frac{1}{2}Z_E^{(1)}\Bigr) = -\frac{\as C_F}{\pi}L_b.
\end{equation}
Both agree with refs.~\cite{Ji:2019ewn,Ebert:2019okf}, where $S_r^{(1)}$ follows from eqs.~(18) and (21) of
ref.~\cite{Ji:2019ewn}. The reduced soft function at one loop thus comes entirely from the Wilson line. It is the difference between the self-interaction of the Wilson line and half that of the rectangular loop, and the
exchange between the quark and the Wilson line does not contribute to it.

The dependence on $\zeta_z$ provides a further check. In eq.~\eqref{ftildedec} it resides in the hard contribution and in
$f^{\mathrm{TMD}}$. Differentiating the coefficient of $\delta(1-x)$ with respect to $L_z = \ln(\zeta_z/\mu^2)$, we obtain
\begin{equation}
\frac{d\tilde{f}^{(1)}}{d\ln\zeta_z} = \asc\bigl(-L_b - L_z + 1\bigr)\delta(1-x) = \frac{1}{2}\bigl[K^{(1)} +
G^{(1)}\bigr]\delta(1-x),
\end{equation}
with $K^{(1)} = -(\as C_F/\pi)L_b$ from $f^{\mathrm{TMD}}$ and $G^{(1)} = (\as C_F/\pi)(1-L_z)$ from $H$, in agreement with
the evolution of the quasi-TMDPDF in $\zeta_z$~\cite[eqs.~(12)--(14)]{Ji:2019ewn}. In the decomposition the Collins-Soper kernel arises from the
rapidity divergence. The coefficient of $\ln\nu$ in $B_n^{(1)}$ is $2\asc L_b = -K^{(1)}$ after renormalization, and the
dependence on $p_+$, which accompanies $\nu$ in the ratio $\nu/p_+$, converts it into the dependence on $\zeta_z$.

The renormalization-group consistency of eq.~\eqref{factintro} can be checked in the same way. We define the anomalous
dimensions by $\gamma_F = d\ln F/d\ln\mu$ for the coefficients of $\delta(1-x)$. The terms with $P_{qq}(x)$ come
from the collinear region and are infrared, and they are identical in $\tilde{f}$ and $f^{\mathrm{TMD}}$, so that they do not
enter the comparison. From the renormalized results above, with $dL_b/d\ln\mu = 2$ and
$dL_z/d\ln\mu = -2$, we obtain at one loop
\begin{equation}
\gamma_{\tilde{f}} = \frac{3\as C_F}{2\pi}, \qquad \gamma_{S_r} = -\frac{2\as C_F}{\pi}, \qquad
\gamma_H = \frac{\as C_F}{\pi}\bigl(L_z-1\bigr), \qquad
\gamma_{f^{\mathrm{TMD}}} = \frac{\as C_F}{\pi}\Bigl(\frac{3}{2}-L_z\Bigr),
\end{equation}
where $\gamma_{f^{\mathrm{TMD}}}$ is evaluated at $\zeta = \zeta_z$. They satisfy
$\gamma_{\tilde{f}} + \frac{1}{2}\gamma_{S_r} = \gamma_H + \gamma_{f^{\mathrm{TMD}}}$. The dependence on $\zeta_z$ cancels
between $H$ and $f^{\mathrm{TMD}}$, and the anomalous dimension of the quasi-TMDPDF is independent of $P^z$. For
$\zeta\neq\zeta_z$, the factor $e^{\frac{1}{2}K\ln(\zeta_z/\zeta)}$ compensates the dependence of $\gamma_{f^{\mathrm{TMD}}}$
on $\zeta$ through $dK/d\ln\mu = -2\Gamma_{\mathrm{cusp}}$, with $\Gamma_{\mathrm{cusp}} = \as C_F/\pi$ at one loop. The
anomalous dimension of $S_r$ arises from eq.~\eqref{DWZE}, that is, from the single segment by which the staple exceeds half
of the rectangular loop.

The dependence on $\nu$ is governed in the same way by
$\gamma^{\nu}_F = d\ln F/d\ln\nu$. From eqs.~\eqref{Bn} and \eqref{Splus}, $\gamma^{\nu}_{B_n} = -K^{(1)}$ and
$\gamma^{\nu}_{S_+} = K^{(1)}$, so that $\gamma^{\nu}_{S_{n\hat{z}}} = \frac{1}{2}\gamma^{\nu}_{S_{n\nb}} = K$ at one loop.

\section{Soft limit of the quasi beam function as an operator} \label{operator}

The soft sector in eq.~\eqref{softsector} is the one-loop expansion of a vacuum matrix element of soft Wilson lines. In this
section we derive the matrix element from the operator, express the reduced soft function through it, and use the operator
form to extend the factorization of the quasi beam function to all orders.

\subsection{Soft Wilson lines} \label{softwilson}

The form of this matrix element follows from the steps that produce the soft function of the lightcone TMDPDF, and we first recall them. Decoupling the soft gluons from the incoming
collinear quark by $\xi_n\to Y_n\xi_n$~\cite{Bauer:2001yt} leaves the collinear quark field and supplies the soft Wilson line
\begin{equation} \label{Yn}
Y_n(x) = P\exp\left[ig\int_{-\infty}^{0}ds\, n\cdot A_s(x+sn)\right],
\end{equation}
which runs from $-\infty$ to $x$ since the quark is incoming~\cite{Chay:2004zn}.

The soft Wilson lines along $\nb$ arise from the momentum conservation imposed by the operator~\cite{Chay:2005rz,Chay:2017bmy}.
The Fourier transform with respect to the longitudinal separation of the quark fields produces a delta function for the large
momentum component $\nb\cdot p$. In soft-collinear effective theory the momentum of a collinear field is divided into a large
label part, of the order of $p$, and a small residual part, of the order of $\lambda p$, which is the size of the momenta of
soft gluons. Momentum conservation holds separately for the two parts, and the delta function splits into one for the label
momenta, which acts on the collinear fields and fixes $x$, and one for the residual momenta, $\delta(\eta-\nb\cdot iD_s)$,
which acts on the soft fields. The covariant derivative appears because the operator is invariant
under soft gauge transformations, and the identity
\begin{equation}
\nb\cdot iD_s = Y_{\nb}\,\nb\cdot i\partial\,Y_{\nb}^{\dagger},
\end{equation}
which follows from $\nb\cdot D_s\,Y_{\nb} = 0$, turns the delta function into $Y_{\nb}\,\delta(\eta-\nb\cdot i\partial)\,
Y_{\nb}^{\dagger}$, with one soft Wilson line at the position of each quark field. The Wilson lines along $\nb$ are what
makes the measurement of the residual momentum gauge invariant.

What happens to the two Wilson lines along $\nb$, one at the position of each quark field, depends on what is measured. Near threshold the residual momentum is
measured, and the delta function remains between the two lines~\cite{Chay:2017bmy}. At generic $x$ it is not measured, and the
integration over $\eta$ sets the longitudinal separation of the two lines to zero. In the collinear PDF the two lines are then
at the same point and cancel. When the transverse momentum is measured, the Fourier transform with respect to $\vec{b}_T$
produces a separate delta function for the transverse momentum of order $\lambda$, which the soft gluons carry as well. The two
lines stay at $\vec{0}_T$ and $\vec{b}_T$ and do not cancel.

\begin{figure}[t]
\centering
\includegraphics[width=0.85\textwidth]{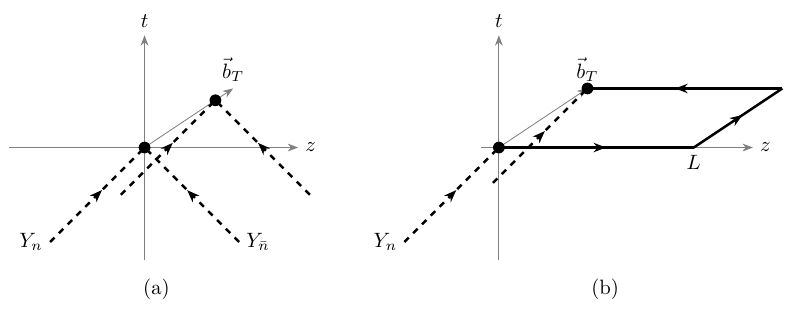}
\caption{\label{figsoft} Wilson lines of the soft functions. (a) The lightcone soft function $S_{n\nb}$. (b) The soft limit
of the quasi beam function, $S_{n\hat{z}}$, in which the quark fields supply lightlike Wilson lines $Y_n$ and the two
ends of the staple are at the same longitudinal position. Lightlike lines are dashed, and spacelike lines are solid. The
lightlike lines extend to infinity, where they are joined by transverse links that are not drawn, since they do not contribute
in covariant gauges.}
\end{figure}

The transverse delta function itself is not converted into Wilson
lines by the equation of motion, since no Wilson line along $\nb$ makes $iD_{\perp}$ a pure derivative. It places the soft
fields at different transverse positions, and gauge invariance requires a transverse link between them, which lies at
infinity and does not contribute in covariant gauges. The soft function, whose Wilson lines are shown in fig.~\ref{figsoft}(a), is
\begin{equation} \label{Snnb}
S_{n\nb}(b_T) = \frac{1}{N_c}\langle 0|\,\mathrm{tr}\,\bigl[Y_n^{\dagger}(b)\,Y_{\nb}(b)\,Y_{\nb}^{\dagger}(0)\,
Y_n(0)\bigr]\,|0\rangle .
\end{equation}
Whether $Y_{\nb}$ runs from $-\infty$ or to $\infty$ depends on the process~\cite{Chay:2004zn}, and for unpolarized TMDs the
soft functions of the Drell-Yan process and of semi-inclusive deep inelastic scattering coincide~\cite{Collins:2004nx}. The
direction matters for spin-dependent TMDs, where it reverses the sign of the Sivers function~\cite{Collins:2002kn}.

The quasi beam function is treated in the same way. The Fourier transform with respect to $b^z$ produces the delta function
for the momentum component along $\hat{z}$, which splits into $\delta(xP^z-\hat{z}\cdot\mathcal{P})$ for the label momenta and
a delta function with $\hat{z}\cdot iD_s$ for the residual momenta. The legs of the staple are the Wilson lines that make
$\hat{z}\cdot iD$ covariant, and they end at $z=L$. Their length is long compared with the soft wavelength $1/k\sim b_T$, so
that their soft part survives with its full length, while the longitudinal separation $b^z\sim 1/P^z$ of the quark fields is
short and the multipole expansion sets $b^z=0$. The transverse delta function again keeps the two legs at $\vec{0}_T$ and
$\vec{b}_T$, and the transverse link at $z=L$ joins them. We denote by $Y_{\hat{z}}(b;L)$ and $Y_{\hat{z}}(0;L)$ the soft parts of the legs
that run from $b$ and from $0$ to $z=L$, and by $Y_T(L;b_T)$ the soft part of the transverse link at $z=L$. They are $W_{\hat{z}}$ and
$W_T$ of section~\ref{quasi} with $A$ replaced by $A_s$. The soft limit of
the quasi beam function, shown in fig.~\ref{figsoft}(b), is then
\begin{equation} \label{Sq}
S_{n\hat{z}}(b_T,L) = \frac{1}{N_c}\langle 0|\,\mathrm{tr}\,\bigl[Y_n^{\dagger}(b)\,Y_{\hat{z}}(b;L)\,Y_T(L;b_T)\,
Y_{\hat{z}}^{\dagger}(0;L)\,Y_n(0)\bigr]\,|0\rangle, \qquad b^{\mu} = (0,\vec{b}_T,0),
\end{equation}
where the points $0$ and $b$, at which the quark fields stood in the quasi beam function, are separated only in the
transverse direction, since the multipole expansion sets $b^z=0$. The
arguments of $S_{n\hat{z}}$ are the transverse separation and the length of the legs, and its dependence on $\mu$ and on the
rapidity regulator is not shown, as for $S_{n\nb}$ in eq.~\eqref{Snnb}. Comparing
eqs.~\eqref{Snnb} and \eqref{Sq}, the lightlike lines $Y_n$ from the quark are kept, and only the pair of lines along $\nb$ is
replaced by the soft part of the staple,
\begin{equation}
Y_{\nb}(b)\,Y_{\nb}^{\dagger}(0)\ \longrightarrow\ Y_{\hat{z}}(b;L)\,Y_T(L;b_T)\,Y_{\hat{z}}^{\dagger}(0;L).
\end{equation}
The same operator appears as the function $\Psi$ of refs.~\cite[eq.~(22)]{Vladimirov:2020ofp},
\cite[eq.~(3.31)]{Rodini:2022wki}, obtained there from the heavy-to-light current. Here it arises as the soft region of the
full-QCD quasi beam function.

\subsection{Reduced soft function in terms of Wilson lines} \label{srwilson}

At one loop, $S_{n\hat{z}}^{(1)} = S_+ + \Delta_W$ by eq.~\eqref{softsector}. With
eqs.~\eqref{softSplus} and \eqref{Sr1}, the reduced soft function can be written as
\begin{equation} \label{Srwilson}
S_r = \frac{S_{n\nb}\,Z_E}{S_{n\hat{z}}^2},
\end{equation}
which at one loop gives $S_r^{(1)} = S^{(1)}_{n\nb} + Z_E^{(1)} - 2S_{n\hat{z}}^{(1)} = -2(\Delta_W - \frac{1}{2}Z_E^{(1)})$.
All three factors in eq.~\eqref{Srwilson} are vacuum expectation values of Wilson lines. The dependence on the rapidity
regulator cancels between $S_{n\nb}$ and $S_{n\hat{z}}^2$ with the symmetric choice of the regulator, and the dependence on $L$
cancels between $Z_E$ and $S_{n\hat{z}}^2$. Equation~\eqref{Srwilson} expresses the soft factor of the factorization formula in terms of the soft limit of the quasi
beam function, without a separate quasi soft function.

In ref.~\cite{Ebert:2022fmh} the same factor is obtained as a ratio of
the naive quasi soft function and the Collins soft function with off-lightcone Wilson lines~\cite[eqs.~(2.36)--(2.38)]{Ebert:2022fmh}. Equation~\eqref{Srwilson} is another ratio for the same object, in which the soft limit of the
quasi beam function takes the place of the off-lightcone soft function, and neither ratio can be computed directly in
Euclidean lattice QCD. In refs.~\cite{Ji:2019sxk,Ji:2019ewn} the rectangular loop removes the self-energy of the Wilson lines in
the conventional way of lattice QCD, and $S_r$ is what remains of the soft function~\cite[eqs.~(3) and (4)]{Ji:2019sxk}. Equation~\eqref{Srwilson} has the same structure. Written as
$S_r=S_{n\nb}\big/\bigl(S_{n\hat{z}}/\sqrt{Z_E}\bigr)^2$, it shows that the rectangular loop removes the self-energy of the
Wilson line from the soft limit of the quasi beam function, and $S_r$ is what remains of the soft function. At one loop
eq.~\eqref{DWZE} locates this remainder in the quasi beam function, as the single straight segment by which the staple
exceeds half of the loop.

Equation~\eqref{Srwilson} follows from the factorization of the quasi beam function into the
hard, collinear and soft sectors, together with $f^{\mathrm{TMD}} = B_n\sqrt{S_{n\nb}}$. Its nontrivial content is that the
right-hand side depends neither on the rapidity regulator nor on $\zeta_z$, which requires the rapidity divergences of $S_{n\hat{z}}$
to be one half of those of $S_{n\nb}$. At one loop this is the folding in eq.~\eqref{fold}, and in section~\ref{allorders} it is shown to
follow at all orders from the absence of rapidity divergences in the quasi beam function.

\subsection{Factorization and rapidity structure to all orders} \label{allorders}

The factorization of the quasi-TMDPDF has been proved to all orders in ref.~\cite[section~3.1]{Ebert:2022fmh} through Lorentz invariance, without an analysis of the momentum regions. The operator form of the soft limit
gives the same factorization from within the quasi beam function, and it identifies the soft factor with a definite part of
it.

The steps leading to the factorization act on the operator and not on individual diagrams. For collinear gluons the
replacement $\rb\cdot A_n\to\frac{1}{2}\nb\cdot A_n$ holds at leading power for any number of gluons, and the collinear sector
of the quasi beam function is the lightcone TMD beam function to all orders. For soft gluons the field redefinition
$\xi_n\to Y_n\xi_n$ decouples them from the quark for any number of gluons, the soft part of the staple keeps its full length
because $L$ is long compared with the soft wavelength, and the multipole expansion sets $b^z=0$. These are the steps that
establish the factorization of the lightcone TMDPDF in soft-collinear effective theory, and here they give $\tilde{B} =
H\,B_n\,S_{n\hat{z}}$ at leading power to all orders in $\alpha_s$, provided that no further region contributes. At one loop the
only other region, with $k_3\lesssim 1/L$ and $k_T\sim 1/b_T$, is suppressed by $b_T^2/L^2$. The same leading regions, with
collinear, soft and hard subdiagrams, appear in the sketch of a proof of eq.~\eqref{factintro} in section~4 of
ref.~\cite{Ji:2019ewn}, where the soft subdiagram describes the radiation between the fast-moving color charge and the staple.
The present analysis makes these regions explicit at one loop and identifies the soft subdiagram with the matrix element
$S_{n\hat{z}}$. The diagrams of higher orders,
including the webs that connect $Y_n$ with the soft part of the staple, are all contained in the matrix element of
eq.~\eqref{Sq} and leave its form unchanged. The one-loop calculation of sections~\ref{regions} and \ref{sectors} confirms the
factorization and identifies the soft sector, but the factorization itself does not rest on it.

The overlap of the collinear and soft sectors is fixed in the same way. The zero-bin contribution of the collinear sector is the limit of
the soft sector in which the gluons become collinear to $n$. In this limit $k_+\gg k_-$, the eikonal propagator of the
longitudinal legs becomes $1/k_3\to 2/k_+$, and the legs act as a line along $\nb$. The zero-bin contribution of the quasi beam function is
then that of the lightcone beam function to all orders, as found at one loop in section~\ref{softzero}, and $B_n$ is the
lightcone beam function with the same subtraction.

The renormalization of $S_{n\hat{z}}$ follows from the factorization by consistency. The ultraviolet divergences of the
staple, namely the linear divergence proportional to its length and the divergences at the cusps and at the endpoints,
renormalize multiplicatively~\cite{Polyakov:1980ca,Dotsenko:1979wb,Ebert:2019tvc}, and the division by $\sqrt{Z_E}$ removes
those associated with the length and the cusps, in agreement with the one-loop result in eq.~\eqref{DWZE}. The anomalous
dimension of $S_{n\hat{z}}$ in $\mu$ is then the difference between that of $\tilde{B}$, which comes from the quark field and the
staple, and those of $H$ and $B_n$.

The rapidity divergences are constrained more strongly, and the argument takes three steps. First, all the Wilson lines in
the operator of the quasi beam function are spacelike and $P^z$ is finite, so that $\tilde{B}$ has no rapidity divergence.
Since $H$ has none either, the rapidity divergences of $B_n$ and $S_{n\hat{z}}$ in $\tilde{B} = H\,B_n\,S_{n\hat{z}}$ cancel
each other. Second, $f^{\mathrm{TMD}} = B_n\sqrt{S_{n\nb}}$ is free of rapidity divergences, so that those of $B_n$ cancel
those of $\sqrt{S_{n\nb}}$. In terms of the rapidity anomalous dimensions the two statements read
\begin{equation}
\gamma^{\nu}_{B_n} + \gamma^{\nu}_{S_{n\hat{z}}} = 0, \qquad \gamma^{\nu}_{B_n} + \frac{1}{2}\gamma^{\nu}_{S_{n\nb}} = 0.
\end{equation}
Third, subtracting one from the other gives
\begin{equation} \label{gammanu}
\gamma^{\nu}_{S_{n\hat{z}}} = \frac{1}{2}\gamma^{\nu}_{S_{n\nb}} = K(b_T,\mu),
\end{equation}
provided that the same rapidity regulator is used for $B_n$, $S_{n\nb}$ and $S_{n\hat{z}}$. The ratio $S_{n\nb}/S_{n\hat{z}}^2$ in
eq.~\eqref{Srwilson} is then free of rapidity divergences, $S_r$ depends only on $b_T$ and $\mu$, and the Collins-Soper kernel
is reproduced by the quasi beam function to all orders. This agrees with the rapidity evolution of $\Psi$ in
refs.~\cite{Vladimirov:2020ofp,Rodini:2022wki} and with ref.~\cite{Ebert:2022fmh}, where the rapidity dependence of the
Collins soft function is governed by the Collins-Soper kernel [eq.~(2.37)], so that the factor relating the naive
quasi-TMDPDF to the TMDPDF separates into the Collins-Soper factor and $g_S^q(b_T,\mu)$, which corresponds to $S_r^{-1/2}$
in eq.~\eqref{factintro} [eqs.~(1.1) and (2.38)], and with the extraction of the kernel from ratios of quasi-TMDPDFs at different momenta, in which the soft factor
cancels~\cite{Ebert:2019tvc}.

The mechanism behind eq.~\eqref{gammanu} can be seen in the diagrams. The rapidity divergences of $S_{n\hat{z}}$ arise when a
soft gluon becomes collinear to $n$, $k_+\to\infty$ in the hemisphere $k_3>0$, where the longitudinal legs act as a line
along $\nb$ as in the zero-bin contribution. The rapidity-divergent subgraphs of $S_{n\hat{z}}$ then coincide with those of $S_{n\nb}$ in
the hemisphere $k_3>0$, and with the symmetry between $n$ and $\nb$ they carry one half of the rapidity divergence of
$S_{n\nb}$. At one loop this is the folding in eq.~\eqref{fold}.

\section{Discussion} \label{discussion}

\subsection{Comparison with quasi soft functions}

In the approaches of refs.~\cite{Ebert:2019okf,Ji:2019ewn}, the soft factor that connects the quasi-TMDPDF with the
TMDPDF is supplied by a vacuum matrix element constructed separately from the quasi beam function. In
ref.~\cite{Ebert:2019okf} the naive quasi soft function $\tilde{S}^q$ is built from Wilson lines along the $z$ direction only,
and the soft factor $\tilde{\Delta}_S^q = 1/\sqrt{\tilde{S}^q}$ is fixed by requiring that all divergences in $L/b_T$ cancel
when it is combined with the quasi beam function~\cite[eqs.~(4.14) and (4.15)]{Ebert:2019okf}. The resulting
quasi-TMDPDF has the single logarithm $\frac{5}{2}L_b$, while the TMDPDF has $\frac{3}{2}L_b$, and the infrared logarithms do not
match~\cite[table~1]{Ebert:2019okf}. Since the soft factors of the TMDPDF always involve two different spatial
directions while the naive quasi soft function involves only the $z$ direction, ref.~\cite{Ebert:2019okf} identified the quasi
soft function as the most likely origin of the mismatch and introduced a bent quasi soft function involving two spatial
directions. Its soft factor carries $-3L_b$ instead of $-2L_b$, and the resulting quasi-TMDPDF has the single logarithm
$\frac{3}{2}L_b$ of the TMDPDF at one loop~\cite[section~4.5]{Ebert:2019okf}. In
refs.~\cite{Ji:2019sxk,Ji:2019ewn} the mismatch is instead accounted for by the reduced soft function $S_r$ in the
factorization formula, which is defined through the soft function with Wilson lines of finite rapidity \cite[eq.~(18)]{Ji:2019ewn},
\cite[eq.~(4)]{Ji:2019sxk} and extracted from a light-meson form factor~\cite[eq.~(19)]{Ji:2019sxk}.

Our decomposition accounts for both observations. The quasi beam function in eq.~\eqref{regionresult} has the single
logarithm $\frac{9}{2}L_b$, in agreement with ref.~\cite{Ebert:2019okf}. Half of the rectangular loop carries $2L_b$ at one
loop, and $\Delta_W$, the self-interaction of the Wilson line in the soft limit, carries $3L_b$. Dividing by either gives
\begin{equation} \label{logledger}
\tfrac{9}{2}L_b - 2L_b = \tfrac{5}{2}L_b \quad \text{(division by $\sqrt{Z_E}$)}, \qquad
\tfrac{9}{2}L_b - 3L_b = \tfrac{3}{2}L_b \quad \text{(division by $\Delta_W$)},
\end{equation}
to be compared with $\frac{3}{2}L_b$ in the TMDPDF. The first is the result with the naive quasi soft function, and its
difference from $\frac{3}{2}L_b$ is $L_b$, which is $-\frac{1}{2}S_r^{(1)}$ in eq.~\eqref{Sr1}. The mismatch found with the naive quasi soft function is thus the
reduced soft function itself, and it originates in the one segment by which the staple exceeds half of the rectangular loop.
In ref.~\cite{Ebert:2019okf} this mismatch was allowed for by a nonperturbative factor $g^q_S$, associated with the failure to
construct a proper quasi soft function~\cite[eq.~(5.3)]{Ebert:2019okf}, and ref.~\cite[eq.~(2.38)]{Ebert:2022fmh} identifies it with
the square root of the reduced soft function.

This is the diagnosis of section~4.5 of ref.~\cite{Ebert:2019okf}, expressed in terms of the regions. There the naive quasi
soft function is split into the exchanges within the $\hat{z}$ and $-\hat{z}$ halves and the exchange between them,
$\tilde{S}^{q(1)} = \tilde{S}^{(1)}_{z/z} + \tilde{S}^{(1)}_{-z/-z} + \tilde{S}^{(1)}_{z/-z}$~\cite[eq.~(4.24)]{Ebert:2019okf}, and it is observed that the quasi beam function at $b^z=0$ contains the same Wilson lines as
half of the quasi soft function, so that $\tilde{S}^{(1)}_{z/z}$ cancels the soft limit of the Wilson-line contribution
to the quasi beam function. In our language this soft limit is the self-interaction of the Wilson line in $S_{n\hat{z}}$, and indeed
$\Delta_W$ in eq.~\eqref{DeltaW} coincides with $\tilde{S}^{(1)}_{z/z}$ in eq.~(4.12) of ref.~\cite{Ebert:2019okf}. The
mismatch was then traced to the exchange $\tilde{S}^{(1)}_{z/-z}$~\cite[eq.~(4.25)]{Ebert:2019okf}. Since
$\tilde{S}^q = Z_E(2L,b_T)$ and $Z_E^{(1)} = 2\Delta_W + \tilde{S}^{(1)}_{z/-z}$, eq.~\eqref{DWZE} gives
\begin{equation}
S_r^{(1)} = -2\Bigl(\Delta_W - \frac{1}{2}Z_E^{(1)}\Bigr) = \tilde{S}^{(1)}_{z/-z} = \asc\Bigl(-\frac{2}{\eps} - 2L_b\Bigr)
\end{equation}
before renormalization, in agreement with eq.~(4.8) of ref.~\cite{Ebert:2019okf}. The exchange between the two halves
of the naive quasi soft function, identified in ref.~\cite{Ebert:2019okf} as the source of the mismatch, is the reduced soft
function at one loop.

Dividing instead by the soft sector without its rapidity-divergent part, $S_{n\hat{z}}/\sqrt{S_{n\nb}}$, which at one loop is
$\Delta_W$ with the single logarithm $3L_b$, gives the TMDPDF multiplied by the hard function. This is the role
played by the bent soft function at one loop. Equation~(4.27) of ref.~\cite{Ebert:2019okf} gives
\begin{equation}
\frac{1}{2}\tilde{S}^{(1)}_{\mathrm{bent}} = \asc\Bigl[\frac{3}{\eps} + 3L_b + 2 + \frac{2\pi L}{b_T}\Bigr] = \Delta_W,
\end{equation}
including the pole, the term linear in $L$ and the constant, which is the second entry in eq.~\eqref{logledger}.

\begin{figure}[t]
\begin{center}
\includegraphics[width=0.95\textwidth]{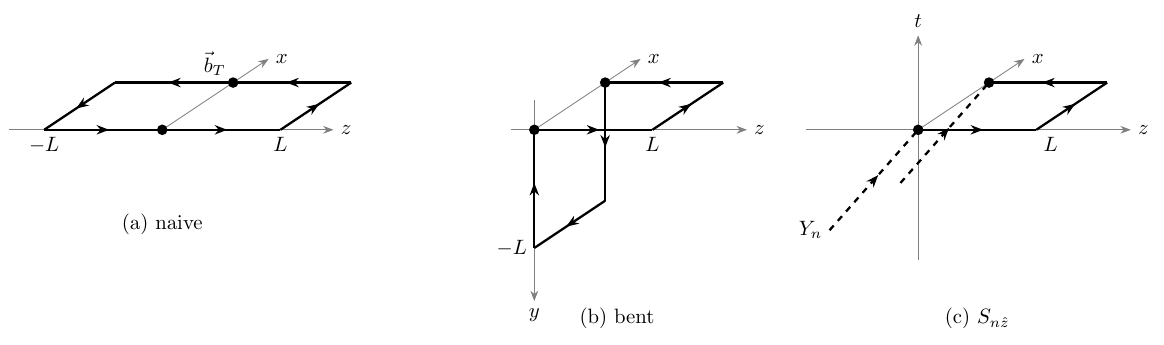}
\end{center}
\caption{Paths of the Wilson lines in (a) the naive quasi soft function $\tilde{S}^q$, (b) the bent quasi soft function
$\tilde{S}_{\mathrm{bent}}$ of ref.~\cite{Ebert:2019okf}, and (c) the soft limit $S_{n\hat{z}}$ of the quasi beam function. The
points $\vec{0}_T$ and $\vec{b}_T$ are marked by dots, $x$ is the direction of $\vec{b}_T$, and $y$ is the direction of
$\bar{n}_\perp$. In (a) the lines along $\hat{z}$ and $-\hat{z}$ join into a rectangular loop of length $2L$. In (b) the
half along $-\hat{z}$ is turned into $-\bar{n}_\perp$, and the contour consists of two staples in perpendicular planes. Each
of them is the Wilson line of $S_{n\hat{z}}$ in (c), in which the lightlike lines along $n$ are dashed.}
\label{figbent}
\end{figure}

The geometry of the bent soft function, shown in fig.~\ref{figbent}, explains this relation. At each of the two points $\vec{0}_T$ and $\vec{b}_T$ it contains one line along $\hat{z}$ and one along
$-\bar{n}_\perp$, both of length $L$, where $\bar{n}_\perp$ is orthogonal to $\hat{z}$ and to $\vec{b}_T$. The lines along
$\hat{z}$ are closed by a transverse link at $-L\hat{z}$, and those along $\bar{n}_\perp$ by a link at $L\bar{n}_\perp$, so
that the contour consists of two staples in perpendicular planes. Since $\hat{z}$, $\bar{n}_\perp$ and $\hat{b}_T$ are
mutually orthogonal, no gluon is exchanged between the two staples at one loop in the Feynman gauge. Each staple then gives
the self-interaction of the Wilson line of the quasi beam function at $b^z=0$, and the four legs, the two links and the exchanges
within each staple add up, for $L\gg b_T$, to
\begin{equation}
\tilde{S}^{(1)}_{\mathrm{bent}} = \asc\Bigl[4\Sigma(L)+2\Sigma(b_T)+4\Bigl(\frac{\pi L}{b_T}-2-2\ln\frac{L}{b_T}\Bigr)\Bigr]
= 2\Delta_W,
\end{equation}
with $\Sigma(\ell)$ the self-energy of a straight line of length $\ell$ in eq.~\eqref{selfen}. The
square root of the bent soft function is one open staple at one loop, in the same way as the square root of the
lightcone soft function is one hemisphere.

The naive quasi soft function, in which the lines from each point run along $\hat{z}$ and
$-\hat{z}$, is the same contour unfolded into a plane~\cite[fig.~9]{Ebert:2019okf}. The two lines at each point join
into one straight line, the points $\vec{0}_T$ and $\vec{b}_T$ lie at the middle of the long sides, and the contour is the
rectangular loop of length $2L$ and width $b_T$, so that $\tilde{S}^q = Z_E(2L,b_T)$. Its square root removes the
dependence on $L$ but misses the contribution of the extra straight segment, which is $S_r$. The bend is what turns the closed
loop into two open staples. Beyond one loop the two staples of the bent soft function are connected by webs with a
three-gluon vertex, and its square root is no longer a single staple, in line with the concern about higher orders expressed in
refs.~\cite{Ebert:2019okf,Ji:2019ewn,Ebert:2022fmh}. The definition of $S_r$ in eq.~\eqref{Srwilson} does not require the two staples to remain unconnected,
since it contains no square root of a closed contour and uses the soft limit $S_{n\hat{z}}$, which contains a single staple by
construction.

The soft functions differ in what happens at the positions of the quark fields, as summarized in table~\ref{tabsoft}. In the lightcone soft function the
line along $n$ from each point meets the line along $\nb$, and this cusp between two lightlike directions is the source of
the rapidity divergence and of the Collins-Soper kernel. The naive quasi soft function is obtained from the lightcone soft
function by replacing both lightlike directions by spacelike ones, $n\to\hat{z}$ and $\nb\to-\hat{z}$~\cite[eqs.~(3.22)--(3.24)]{Ebert:2019okf}. In the quasi beam function the boost argument of LaMET replaces only one lightlike direction, the
direction $\nb$ of the Wilson line, by $\hat{z}$, and the naive quasi soft function applies the same argument to both. The two lines at each point then run in opposite directions and join into a straight line, so that the
cusp disappears and the rapidity dependence goes with it. The line along $n$, however, comes from the state and not from
the operator, and since the boost brings $\hat{z}$ close to $\nb$, it cannot take the place of $n$. In $S_{n\hat{z}}$ only the direction $\nb$ is replaced by the
direction $\hat{z}$ of the legs, while the lightlike line of the quark is kept. The cusp at each point joins the lightlike
direction $n$ to the spacelike direction $\hat{z}$, and it carries one half of the rapidity divergence of the lightlike cusp,
as the folding in eq.~\eqref{fold} shows at one loop and eq.~\eqref{gammanu} at all orders. The bent soft function has at each point a right-angled cusp between two
spacelike directions, which carries no rapidity divergence, and at one loop its two staples reproduce $\Delta_W$ because they
do not interact.

\begin{table}[t]
\centering
\small
\begin{tabular}{lllcl}
\hline
Soft function & Lines at each point & Cusp at each point & Rapidity div. & One loop \\
\hline
$S_{n\nb}$ & $n$ and $\nb$, to infinity & lightlike--lightlike & full & $\frac{1}{2}S_{n\nb}^{(1)} = S_+$ \\
$\tilde{S}^q$ (naive) & $\hat{z}$ and $-\hat{z}$, length $L$ & none (straight line) & none & $\tilde{S}^q = Z_E(2L,b_T)$ \\
$\tilde{S}_{\mathrm{bent}}$ & $\hat{z}$ and $-\bar{n}_\perp$, length $L$ & spacelike, right angle & none &
$\frac{1}{2}\tilde{S}^{(1)}_{\mathrm{bent}} = \Delta_W$ \\
$S_{n\hat{z}}$ & $n$, and $\hat{z}$ of length $L$ & lightlike--spacelike & one half & $S_{n\hat{z}}^{(1)} = S_+ + \Delta_W$ \\
\hline
\end{tabular}
\caption{\label{tabsoft} Wilson-line content of the soft functions compared in this section. The lines at each of the
points $\vec{0}_T$ and $\vec{b}_T$ and the cusp they form there determine the rapidity divergence. The last column gives the
one-loop relation to the quantities of sections~\ref{sectors} and \ref{softfactor}.}
\end{table}

The calculation of sections~\ref{regions} to \ref{operator} organizes the same physics differently. The soft input is
not constructed but obtained, as the soft limit of the quasi beam function, in which the quark supplies a lightlike Wilson line
along $n$ through the decoupling of the soft gluons and the Wilson line is kept intact. The resulting matrix element $S_{n\hat{z}}$ contains only one lightlike direction, while the
lightcone soft function $S_{n\nb}$ connects the two directions $n$ and $\nb$. The two are related because the TMDPDF
contains only the square root of $S_{n\nb}$. With the symmetric regulator, $\sqrt{S_{n\nb}}$ is the contribution of one
hemisphere, and the folding in eq.~\eqref{fold} shows that the exchange between the lightlike line and the Wilson line, integrated
over all $k_3$, reproduces exactly this hemisphere. The folding works because the eikonal propagator $1/k_3$ of the legs changes sign under the
interchange of $n$ and $\nb$, which maps the boundary $k_3=0$ between the hemispheres onto itself. The direction $\nb$ enters
our calculation only through the zero-bin contribution, whose integrand is that of $S_{n\nb}$, and through the comparison with
$f^{\mathrm{TMD}}$. The quasi beam function carries its own second direction, the direction $\hat{z}$ of the Wilson line, and
its soft limit provides what the square root of the two-directional soft function provides in the TMDPDF.

Since $S_{n\hat{z}}$ contains a lightlike Wilson line, it is not computable in Euclidean lattice QCD, and
eq.~\eqref{Srwilson} is not a proposal for a lattice determination of $S_r$. Its role is to identify which part of
the quasi beam function supplies the soft factor and to give $S_r$ a definition in terms of Wilson lines, which can be used in
perturbation theory and compared with the definition in eq.~(18) of ref.~\cite{Ji:2019ewn}.

\subsection{Role of the staple length} \label{lrole}

In ref.~\cite{Ebert:2019okf} the finite length is regarded as the regulator of the analog of rapidity divergences, and the term
linear in $L/b_T$ as its manifestation. For lightlike Wilson lines of finite length this is the case, since the phase
$e^{in\cdot kL}$ regulates the limit $n\cdot k\to 0$. For the spacelike legs of the staple the corresponding singularity at
$k_3\to 0$ lies at central rapidity. The term linear in $L$ appears also in the rectangular Wilson loop, which has no lightlike
direction, and it is the static potential between the antiparallel legs. The rapidity divergences appear only in the
collinear and soft sectors, where they cancel in the sum, and their remnant in the quasi beam function is the dependence on
$P^z$ through the logarithms $L_z$ that multiply the Collins-Soper kernel.

Nor can the length be traded for a rapidity regulator. If the limit $L\to\infty$ is taken before the integration, the
eikonal factors of a leg combine into the pinched product $1/[(k_3-i0)(k_3+i0)]$, which is singular at $k_3=0$. There an
on-shell gluon has $k_+=k_-=k_T$, at central rapidity, and a regulator acting on $k_+$ leaves the singularity untouched. A
finite length is required for the individual diagrams to be well defined, and the rapidity regulator enters only through the
division into regions.

\subsection{Beyond one loop}

The operator form of the soft limit settles the factorization of the quasi beam function and the rapidity structure of
$S_{n\hat{z}}$ to all orders, as shown in section~\ref{allorders}. Two observations support eq.~\eqref{gammanu} from a
different angle. In the derivation of the Collins-Soper
equation~\cite{Collins:2011zzd}, the dependence of a soft factor on the rapidity of one of its Wilson lines is governed by
$K(b_T,\mu)$ once that rapidity is large, independently of the directions of the other lines. In $S_{n\hat{z}}$ only the line along $n$
has an unbounded rapidity, while the direction of the longitudinal legs is fixed, and the rapidity dependence of $S_{n\hat{z}}$ is that of a single
lightlike line, one half of that of $S_{n\nb}$, as eq.~\eqref{gammanu} requires.
The second observation concerns the configurations in which the gluons attached to the Wilson line lie in opposite hemispheres. They are of the type that produces non-global logarithms when two regions are
treated differently~\cite{Dasgupta:2001sh}, and here the two regions are the hemispheres. In the evolution of non-global
observables the global part is governed by the universal cusp anomalous dimension, and the correlated emission into the
two regions adds a separate non-global contribution~\cite{Banfi:2002hw}. By analogy, these configurations are not expected
to modify the coefficient of the rapidity divergence, whose dependence on $\mu$ is fixed by the cusp anomalous dimension,
$dK/d\ln\mu = -2\Gamma_{\mathrm{cusp}}$, in accordance with eq.~\eqref{gammanu}.

What the consistency does not fix is the part of $S_{n\hat{z}}$ free of rapidity divergences. At two loops, webs whose gluons
lie in both hemispheres produce such terms, and they enter $S_r$. Whether $S_{n\hat{z}}$ can be replaced by
$\sqrt{S_{n\nb}Z_E}$, or by the bent soft function in fig.~\ref{figbent}, is a question about these webs and not about the
factorization. An explicit calculation at two loops answers it and tests eq.~\eqref{gammanu} at the same time. It requires the complete set of webs of order $C_FC_A$ and $C_FT_Fn_f$ with at least one gluon
attached to a lightlike line, since the webs attached to the lightlike lines alone vanish in the Feynman gauge. The relation
does not hold web by web. The crossed pair of exchanges, one between the line along $n$ at the origin and the leg at
$\vec{b}_T$ and the other between the line along $n$ at $\vec{b}_T$ and the leg at the origin, is the product of two one-loop
exchanges with the color factor $-C_FC_A/2$. Each factor folds as at one loop, and the web of $S_{n\hat{z}}$ is one quarter of the
corresponding web of $S_{n\nb}$. It contains double poles in $\eta$, which must cancel among the webs of each function, and
the comparison of $S_{n\hat{z}}$ with $S_{n\nb}^{1/2}$ is meaningful only for the complete set. In the same set, the regulator has to be
applied to each gluon attached to a lightlike line, with the momentum it carries into the line and the hemisphere fixed in the
frame defined by $n$ and $\nb$, which extends the regulator of ref.~\cite{Chay:2020jzn} to several gluons. The two-loop
rapidity anomalous dimension with the $\eta$ regulator~\cite{Lubbert:2016rku} and the two-loop
quasi-TMDPDF~\cite{delRio:2023pse} provide the reference values for $K$ and for $S_r$. We leave this calculation to a separate
publication.

\section{Conclusions} \label{conclusion}

We have computed the one-loop quasi beam function for a staple-shaped Wilson line of finite length $L$ with
momentum-space Feynman rules, evaluating each diagram separately. The result is exact in $p^zb_T$ and $b_T/L$, the
logarithms of $L$ cancel identically, and the remaining linear dependence on $L$ is removed by the rectangular Wilson loop of
length $2L$. At leading power the result agrees with ref.~\cite{Ebert:2019okf}, with the self-energy of the transverse link
supplying $1/\eps+L_b$ to the coefficient of $\delta(1-x)$.

The decomposition by the method of regions assigns each diagram to definite regions. The hard contribution comes entirely
from the virtual contribution of the sail diagrams and gives the matching coefficient $H$. The collinear region, after the zero-bin
subtraction, is the lightcone TMD beam function, since collinear gluons see the longitudinal legs as a lightlike line along $\nb$. The soft
region, the soft limit of the quasi beam function, consists of the exchange between a lightlike line along $n$ and the Wilson line
and of the self-interaction of the Wilson line. The collinear and soft sectors each contain an ultraviolet rapidity divergence, and with the hemisphere
$\eta$ regulator the two cancel between the sectors. The regulator has to be symmetric between the two hemispheres,
because the eikonal propagator of the longitudinal legs is singular on their boundary.

In the soft function of the TMDPDF a soft gluon is exchanged between Wilson lines along $n$ and $\nb$, whereas in the soft
region of the quasi beam function it is exchanged between the lightlike line along $n$ and the staple along $\hat{z}$. We have
shown that the two are related by the folding of the hemispheres, in which the interchange of $k_+$ and $k_-$ maps the
hemisphere $k_3<0$ onto $k_3>0$ and the two are added. Integrated over the longitudinal momentum of the gluon, the exchange
between the lightlike line and the Wilson line is then equal to one hemisphere of the lightcone soft function, which is its
square root at one loop. The quasi beam function thus reproduces the soft physics of the TMDPDF without a separately
constructed soft function, the direction $\hat{z}$ of its Wilson line playing the role of the second lightlike direction. The self-interaction of the Wilson line, compared with half of the
rectangular Wilson loop, leaves the contribution of one straight segment, which is the reduced soft function
$S_r^{(1)} = -(\as C_F/\pi)L_b$. This leads to the expression $S_r = S_{n\nb}Z_E/S_{n\hat{z}}^2$ in terms of vacuum matrix elements of
Wilson lines. The same decomposition shows that the mismatch found with the simplest quasi soft function is $S_r$ itself, and
that half of the bent quasi soft function of ref.~\cite{Ebert:2019okf} equals the self-interaction of the Wilson line at one loop.

The soft limit has a definite operator form. Soft gluons decouple from the quark through the lightlike Wilson lines $Y_n$, and
the Wilson lines that dress the momentum operator fixing $x$ survive as soft Wilson lines, because the measured transverse
momentum keeps them at different transverse positions. The same steps give $Y_{\nb}$ in the lightcone soft function and
$Y_{\hat{z}}$, ending at $z=L$, in $S_{n\hat{z}}$, so that the two soft functions differ only in the line that dresses the
momentum operator.

Since these steps act on operators, the factorization $\tilde{B} = H\,B_n\,S_{n\hat{z}}$ holds at leading power to all orders,
in agreement with the proof of ref.~\cite{Ebert:2022fmh}, which our analysis complements by the momentum regions.
The quasi beam function has no rapidity divergence, and the rapidity divergences of $S_{n\hat{z}}$ must cancel those of $B_n$.
The rapidity anomalous dimension of $S_{n\hat{z}}$ is then one half of that of $S_{n\nb}$, and the Collins-Soper kernel is
reproduced to all orders. The parts of $S_{n\hat{z}}$ free of rapidity divergences are not fixed by this argument and enter
$S_r$. Whether $S_r$ can be expressed through $S_{n\nb}$ and $Z_E$, or through the bent soft function, beyond one loop is
decided by the webs that connect the lightlike line with the Wilson line, and an explicit calculation at two loops is left to a
separate publication. The finite length $L$, finally, is not a regulator of rapidity divergences in the quasi beam function. It
controls the self-interaction of the Wilson line, which the division by $\sqrt{Z_E}$ removes, and it is required for the
individual diagrams to be well defined.

\appendix

\section{Feynman rules for the staple-shaped Wilson line} \label{app:feyn}

\subsection{Wilson lines in momentum space}

We derive the Feynman rules in momentum space. Let $\rb^{\mu} = (0, \vec{0}_T, -1)$ be the unit vector along the $z$
direction, with the sign chosen such that $\rb\cdot p = p_3$ for $p^{\mu} = (p_0, \vec{p}_T, p_3)$. In terms of the unit
vector $\hat{z}^{\mu} = (0,\vec{0}_T,1)$ it is $\rb^{\mu} = -\hat{z}^{\mu}$. The
Wilson lines along $z$ extending to $\mp\infty$ are written as
\begin{align}
W_3 (x_3) &= P\exp\left[ ig\int_{-\infty}^{x_3} dz_3\, \rb\cdot A(z_3)\right] = \exp\left[ \frac{1}{\rb\cdot\mathcal{P}+i0}
\bigl(-g\rb\cdot A\bigr)\right], \nonumber \\
\widetilde{W}_3 (x_3) &= \overline{P}\exp\left[ -ig\int_{x_3}^{\infty} dz_3\, \rb\cdot A(z_3)\right] = \exp\left[
\frac{1}{\rb\cdot\mathcal{P}-i0}\bigl(-g\rb\cdot A\bigr)\right],
\end{align}
where $\mathcal{P}$ is the momentum operator acting on the gauge field, together with their Hermitian conjugates. A finite
segment between $x_3$ and $L$ is obtained as the product of two lines with a common endpoint at infinity,
\begin{equation}
[x_3, L] = \widetilde{W}_3(x_3)\widetilde{W}_3^{\dagger}(L) = W_3(x_3) W_3^{\dagger}(L),
\end{equation}
and both representations give the same segment. The bracket $[x,y]$ is the Wilson line from $y$ to $x$, which transforms as
$U(x)\,[x,y]\,U^{\dagger}(y)$ under gauge transformations. In the following we use $\widetilde{W}$. The transverse Wilson lines
along the unit vector $\bh$ are defined in the same way with $\rb\cdot A$ replaced by $\bh\cdot\vec{A}_T$ and
$\rb\cdot\mathcal{P}$ by $\bh\cdot\vec{\mathcal{P}}_T$. The operator in eq.~\eqref{Bcoord} is then
\begin{equation}
\overline{\psi}(b_z,\vec{b}_T)\,\frac{\gamma^z}{2}\, [b_z,L]_{\vec{b}_T}\, [\vec{b}_T,\vec{0}_T]_L\,
[L,0]_{\vec{0}_T}\, \psi(0,\vec{0}_T),
\end{equation}
where the three brackets are the longitudinal leg at $\vec{b}_T$, the transverse link at $z=L$, and the longitudinal
leg at the origin, respectively. The subscripts give the transverse position of the legs and the longitudinal position of the link.

\subsection{One-gluon vertices}

\begin{figure}[t]
\begin{center}
\includegraphics[width=\textwidth]{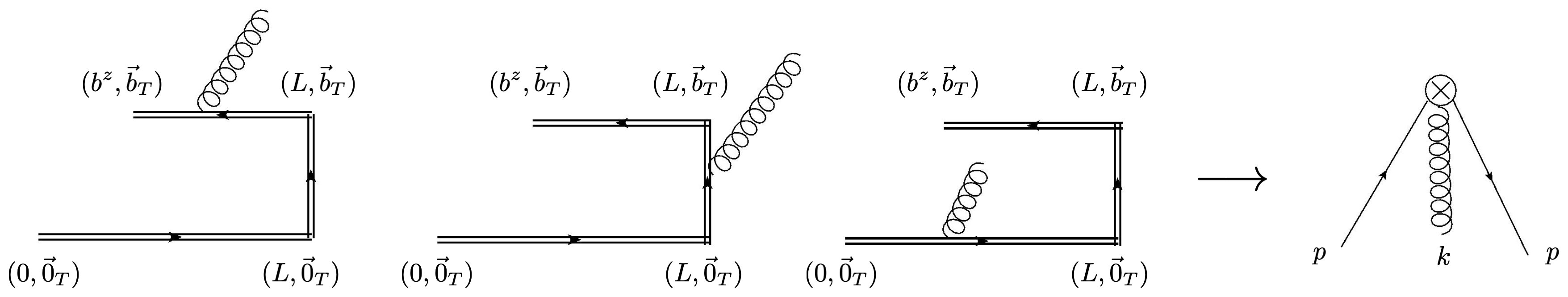}
\end{center}
\caption{One-gluon vertex of the staple-shaped Wilson line. The gluon is attached to the leg at $\vec{b}_T$, to the
transverse link at $z=L$, or to the leg at $\vec{0}_T$, and the sum of the three attachments is the vertex $\otimes$ in
eq.~\eqref{onegluon}, with the gluon momentum $k$ flowing into it.}
\label{figfeyn1}
\end{figure}

Expanding each segment to first order in $g$, with the gluon momentum $k$ flowing into the vertex, the three segments give
\begin{align}
[b_z, L]_{\vec{b}_T}&: \quad g\,\frac{\rb^{\mu}T^a}{k_3-i0}\, e^{i\vec{k}_T\cdot\vec{b}_T}
\bigl(e^{ik_3L} - e^{ik_3b_z}\bigr), \nonumber \\
[\vec{b}_T,\vec{0}_T]_L&: \quad g\,\frac{\bh^{\mu}T^a}{\bh\cdot\vec{k}_T-i0}\, e^{ik_3L}
\bigl(1-e^{i\vec{k}_T\cdot\vec{b}_T}\bigr), \nonumber \\
[L, 0]_{\vec{0}_T}&: \quad g\,\frac{\rb^{\mu}T^a}{k_3-i0} \bigl(1 - e^{ik_3L}\bigr).
\end{align}
For the leg at $\vec{b}_T$, the two factors $\widetilde{W}_3(b_z)$ and $\widetilde{W}_3^{\dagger}(L)$ give
$-g\rb^{\mu}T^a e^{ik_3b_z+i\vec{k}_T\cdot\vec{b}_T}/(k_3-i0)$ and
$-g\rb^{\mu}T^a e^{ik_3L+i\vec{k}_T\cdot\vec{b}_T}/(-k_3+i0)$, respectively, and their sum is the first line. The
others follow in the same way. Adding the three, the one-gluon vertex of the Wilson line for an incoming gluon is
\begin{equation} \label{onegluon}
g\, T^a \left\{ \frac{\rb^{\mu}}{k_3-i0}\Bigl[ \bigl(1-e^{ik_3L}\bigr) + e^{i\vec{k}_T\cdot\vec{b}_T}
\bigl(e^{ik_3L}-e^{ik_3b_z}\bigr)\Bigr] + \frac{\bh^{\mu}}{\bh\cdot\vec{k}_T-i0}\, e^{ik_3L}
\bigl(1-e^{i\vec{k}_T\cdot\vec{b}_T}\bigr) \right\}.
\end{equation}
For an outgoing gluon, $k$ is replaced by $-k$. The three attachments and the resulting vertex are shown in
fig.~\ref{figfeyn1}.

\subsection{Two-gluon vertices}

At second order in $g$, the two gluons, with momenta $k$ and $q$, Lorentz indices $\mu$ and $\nu$ and color indices $a$
and $b$, can be attached to the same segment or to different segments. For two gluons on the same longitudinal segment,
the second-order term of a single line and the product of the first-order terms of the two lines forming the segment
contribute. For the leg at the origin, for instance, the sum of the three contributions is
\begin{align}
g^2\rb^{\mu}\rb^{\nu}&\left\{ T^aT^b\left[ \frac{1}{q_3k_3}\bigl(1-e^{iq_3L}\bigr) + \frac{1}{(k_3+q_3)k_3}\bigl(1-e^{i(k_3+q_3)L}
\bigr)\right] \right. \nonumber \\
&\left. + T^bT^a\left[ \frac{1}{k_3q_3}\bigl(1-e^{ik_3L}\bigr) + \frac{1}{(k_3+q_3)q_3}\bigl(1-e^{i(k_3+q_3)L}\bigr)
\right]\right\},
\end{align}
with the prescriptions $k_3-i0$, $q_3-i0$ and $k_3+q_3-i0$ understood. The leg at $\vec{b}_T$ and the transverse
link give analogous expressions. Since $\bh\cdot\rb=0$, the gluons attached to the longitudinal legs and to the transverse
link are not contracted with each other in the Feynman gauge, and the only mixed contributions needed are those with one
gluon on each longitudinal leg. For these, the four orderings give
\begin{align}
g^2\frac{\rb^{\mu}\rb^{\nu}}{(k_3-i0)(q_3-i0)} \Bigl[ &T^aT^b e^{i\vec{k}_T\cdot\vec{b}_T}\bigl(e^{ik_3b_z}
-e^{ik_3L}\bigr)\bigl(e^{iq_3L}-1\bigr) \nonumber \\
&+ T^bT^a e^{i\vec{q}_T\cdot\vec{b}_T}\bigl(e^{iq_3b_z}-e^{iq_3L}\bigr)\bigl(e^{ik_3L}-1\bigr)\Bigr].
\end{align}

\begin{figure}[t]
\begin{center}
\includegraphics[width=\textwidth]{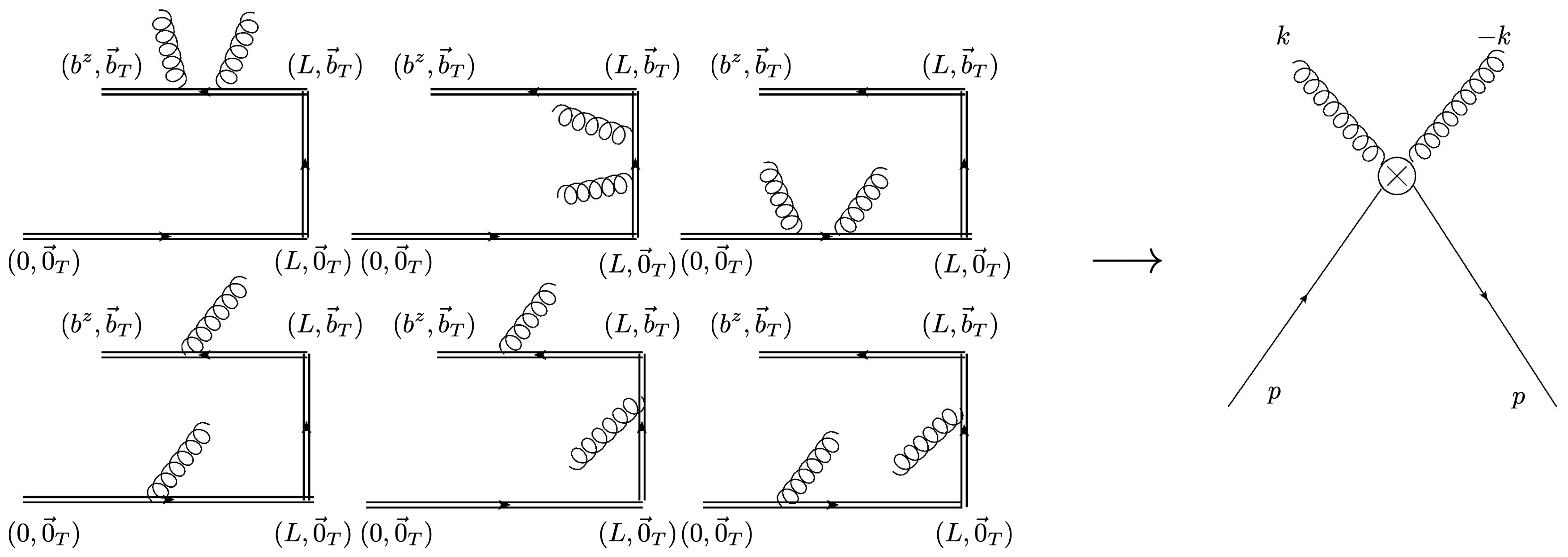}
\end{center}
\caption{Two-gluon vertex of the staple-shaped Wilson line. Upper row: both gluons on the leg at $\vec{b}_T$, on the
transverse link, and on the leg at $\vec{0}_T$. Lower row: one gluon on each longitudinal leg, and one gluon on a
longitudinal leg and the other on the transverse link. When the two gluons are contracted with each other, with momenta $k$
and $-k$, the sum is the vertex in eq.~\eqref{twogluon}. The attachments with one gluon on a leg and the other on the
transverse link do not contribute in the Feynman gauge, since $\rb\cdot\bh=0$. Closing the vertex with a propagator gives
fig.~\ref{figdiag}(c), which corresponds to the Wilson-line self-energy, or tadpole, diagram of ref.~\cite{Ebert:2019okf}.}
\label{figfeyn}
\end{figure}

When the two gluons are contracted with each other, one gluon enters the vertex with momentum $k$ and the other leaves it with
the same momentum, so that $q=-k$. Setting $q=-k$ directly turns the terms with $1/(k_3+q_3)$ into $0/0$. We
therefore keep $q$ general and take the limit $q\to -k$ at the end, where
\begin{equation}
\lim_{q_3\to -k_3}\frac{1-e^{i(k_3+q_3)L}}{k_3+q_3} = -iL,
\end{equation}
and these limits produce the terms linear in the lengths of the segments. Collecting all contributions, the two-gluon
vertex closed by a single propagator takes the form
\begin{equation} \label{twogluon}
g^2\bigl(T^aT^b A^{\mu\nu} + T^bT^a A^{\mu\nu *}\bigr),
\end{equation}
with
\begin{align}
A^{\mu\nu} = {}&\rb^{\mu}\rb^{\nu}\left\{ -\frac{1}{(k_3-i0)(k_3+i0)}\Bigl[ 2-e^{-ik_3L}\bigl(1+e^{ik_3b_z}\bigr) \right.
\nonumber \\
&\qquad\left. + e^{i\vec{k}_T\cdot\vec{b}_T}\bigl(e^{ik_3b_z}-e^{ik_3L}\bigr)\bigl(e^{-ik_3L}-1\bigr)\Bigr]
- \frac{i(b_z-L)}{k_3+i0} - \frac{iL}{k_3-i0}\right\} \nonumber \\
&+ \bh^{\mu}\bh^{\nu}\left[ -\frac{1-e^{i\vec{k}_T\cdot\vec{b}_T}}{(k_b-i0)(k_b+i0)}
- \frac{i\,b_T}{k_b+i0}\right], \qquad k_b \equiv \bh\cdot\vec{k}_T.
\end{align}
The term with $T^bT^a$ is the complex conjugate of the term with $T^aT^b$, as follows from interchanging the roles of the
two gluons. Since the vertex is symmetric in the two gluons, the contraction with a single propagator carries the symmetry
factor $1/2$, and for the color factor $C_F$ the combination becomes $C_F\,\mathrm{Re}A^{\mu\nu}$. The terms linear in
$L$, $b_z$ and $b_T$ do not contribute. Once the length is factored out, their integrands contain no phase and no other
scale, and they vanish in dimensional regularization. The attachments contained in
eq.~\eqref{twogluon} are shown in fig.~\ref{figfeyn}.

\subsection{Equivalence with the coordinate-space rules}

The same vertices follow from the coordinate-space expression used in ref.~\cite{Ebert:2019okf}. Parametrizing the three
segments as $\gamma_1(s) = (0,\vec{0}_T, Ls)$, $\gamma_2(s) = (0, s\vec{b}_T, L)$ and
$\gamma_3(s) = (0,\vec{b}_T, L+s(b_z-L))$ with $0\le s\le 1$, the one-gluon term is
$igT^a\int_0^1 ds\,\gamma'^{\mu}(s)e^{-ik\cdot\gamma(s)}$. For $\gamma_1$, with $\gamma_1'^{\mu} = L\hat{z}^{\mu} =
-L\rb^{\mu}$ and $k\cdot\gamma_1(s) = -k_3Ls$, this gives
\begin{equation}
-igT^a\, L\,\rb^{\mu} \int_0^1 ds\, e^{ik_3Ls} = g\frac{\rb^{\mu}T^a}{k_3}\bigl(1-e^{ik_3L}\bigr),
\end{equation}
and $\gamma_2$ and $\gamma_3$ reproduce the other two terms of eq.~\eqref{onegluon}, apart from the $i0$ prescriptions,
which are fixed by the representation in terms of lines extending to infinity. At second order, the unordered double
integrals $(igT^a)(igT^b)\int_0^1ds\int_0^1dt\,\gamma_i'^{\mu}(s)\gamma_j'^{\nu}(t)\,e^{-ik\cdot[\gamma_i(s)-\gamma_j(t)]}$ give,
as the coefficients of $g^2T^aT^b\rb^{\mu}\rb^{\nu}$ for the legs and of $g^2T^aT^b\bh^{\mu}\bh^{\nu}$ for the link,
\begin{align}
(i,j)=(1,1)&: \quad -\frac{1}{k_3^2}\bigl(2-e^{ik_3L}-e^{-ik_3L}\bigr), \nonumber \\
(i,j)=(3,3)&: \quad -\frac{1}{k_3^2}\bigl(2-e^{ik_3(b_z-L)}-e^{-ik_3(b_z-L)}\bigr), \nonumber \\
(i,j)=(2,2)&: \quad -\frac{b_T^2}{(\vec{k}_T\cdot\vec{b}_T)^2}\bigl(2-e^{i\vec{k}_T\cdot\vec{b}_T}
-e^{-i\vec{k}_T\cdot\vec{b}_T}\bigr), \nonumber \\
(i,j)=(1,3)&: \quad -\frac{1}{k_3^2}e^{-i\vec{k}_T\cdot\vec{b}_T}\bigl(e^{-ik_3b_z}-e^{-ik_3L}\bigr)
\bigl(e^{ik_3L}-1\bigr),
\end{align}
and $(3,1)$ is the complex conjugate of $(1,3)$. Since $k$ enters at $\gamma_i(s)$ and leaves at $\gamma_j(t)$, the opposite of the assignment in eq.~\eqref{twogluon},
these reproduce after $k\to -k$ the corresponding terms in
eq.~\eqref{twogluon} apart from the $i0$ prescriptions. The path ordering, which the unordered integrals do not contain, generates the imaginary terms linear
in the lengths, which vanish in dimensional regularization.

\section{Contributions of the individual diagrams} \label{app:k0}

\subsection{Notation}

After the integration over $k^0$, the remaining integrals are over $k_3$ and $\vec{k}_T$. We write the transverse
integration as
\begin{equation}
\int[dk_T] \equiv 4\pi\Bigl(\frac{\mu^2e^{\gamma_E}}{4\pi}\Bigr)^{\eps}\int\frac{d^{2-2\eps}k_T}{(2\pi)^{2-2\eps}},
\end{equation}
so that for a function of $k_T^2$ alone
\begin{equation}
\int[dk_T]\, f(k_T^2) = \frac{(\mu^2e^{\gamma_E})^{\eps}}{\Gamma(1-\eps)}\int_0^{\infty}dk_T^2\,
(k_T^2)^{-\eps}f(k_T^2).
\end{equation}
We use $E = (k_3^2+k_T^2)^{1/2}$ and $E_q = [(p-k_3)^2+k_T^2]^{1/2}$, and the terms $h_j$ of the sum of the two
residues in eqs.~\eqref{k0sail} and \eqref{hj}. We label the virtual and the real contributions by $V$ and $R$, with the superscripts
$a$, $b$ and $c$ referring to the diagrams in fig.~\ref{figdiag}. All expressions include the complex-conjugate diagrams,
and only their real parts are to be taken.

\subsection{Sail diagrams}

The virtual contributions of the sail diagrams are
\begin{align}
V^a_j &= -\asc\,\delta\bigl(p(1-x)\bigr)\int_{-\infty}^{\infty}dk_3\int[dk_T]\,\frac{1-e^{ik_3L}}{k_3}\, h_j,
\nonumber \\
V^a_{j+6} &= -\asc\,\delta\bigl(p(1-x)\bigr)\int_{-\infty}^{\infty}dk_3\int[dk_T]\,
\frac{e^{ik_3L}e^{i\vec{k}_T\cdot\vec{b}_T}}{k_3-i0}\, h_j, \qquad j=1,\dots,4,
\nonumber \\
V^a_5 &= -\frac{1}{2}\asc\,\delta\bigl(p(1-x)\bigr)\int_{-\infty}^{\infty}dk_3\int[dk_T]\,
e^{ik_3L}\bigl(1-e^{i\vec{k}_T\cdot\vec{b}_T}\bigr)\frac{k_3}{k_T^2\,E},
\nonumber \\
V^a_6 &= -\frac{1}{2}\asc\,\delta\bigl(p(1-x)\bigr)\int_{-\infty}^{\infty}dk_3\int[dk_T]\,
e^{ik_3L}\bigl(1-e^{i\vec{k}_T\cdot\vec{b}_T}\bigr)\frac{p-k_3}{k_T^2\,E_q}.
\end{align}
$V^a_1$ to $V^a_4$ come from the leg at the origin, $V^a_5$ and $V^a_6$ from the transverse link, and $V^a_7$ to $V^a_{10}$
from the term with $e^{ik_3L}$ of the leg at $\vec{b}_T$. In $V^a_1$ to $V^a_4$ the factor $(1-e^{ik_3L})/k_3$ is regular at
$k_3=0$, and the prescription $k_3-i0$ can be added to it. The terms with $h_j$ combine as
\begin{equation}
\sum_{j=1}^{4}\bigl(V^a_j+V^a_{j+6}\bigr) = -\asc\,\delta\bigl(p(1-x)\bigr)\int_{-\infty}^{\infty}dk_3\int[dk_T]
\left[\frac{1}{k_3-i0} + \frac{e^{ik_3L}\bigl(e^{i\vec{k}_T\cdot\vec{b}_T}-1\bigr)}{k_3-i0}\right] h ,
\end{equation}
which is the identity in eq.~\eqref{splitid}. The two contributions of the virtual sail in section~\ref{quasi} are
\begin{align}
\tilde{B}^{(1)}_{\mathrm{sail,v1}} &= -\asc\,\delta\bigl(p(1-x)\bigr)\int_{-\infty}^{\infty}\frac{dk_3}{k_3-i0}\int[dk_T]\, h ,
\nonumber \\
\tilde{B}^{(1)}_{\mathrm{sail,v2}} &= -\asc\,\delta\bigl(p(1-x)\bigr)\int_{-\infty}^{\infty}dk_3\int[dk_T]\,
\frac{e^{ik_3L}\bigl(e^{i\vec{k}_T\cdot\vec{b}_T}-1\bigr)}{k_3-i0}\, h + V^a_5 + V^a_6 ,
\end{align}
so that $\sum_{i=1}^{10}V^a_i = \tilde{B}^{(1)}_{\mathrm{sail,v1}}+\tilde{B}^{(1)}_{\mathrm{sail,v2}}$. The first depends
on $p$ alone, and the second contains all the dependence of the virtual sail on $L$ and $b_T$.

The real contributions come from the term with $e^{ik_3b^z}$ of the leg at $\vec{b}_T$,
\begin{equation}
R^a_j = \asc\int[dk_T]\, e^{i\vec{k}_T\cdot\vec{b}_T}\,\PV\frac{1}{k_3}\, h_j\Big|_{k_3=p(1-x)}, \qquad j=1,\dots,4,
\end{equation}
where the sail diagram and its complex conjugate combine into the principal value. Explicitly,
\begin{align}
R^a_1 &= \asc\int[dk_T]\, e^{i\vec{k}_T\cdot\vec{b}_T}\,\frac{p\,x}{k_T^2}\,
\frac{1}{\sqrt{k_T^2+p^2(1-x)^2}}, \nonumber \\
R^a_2 &= \asc\,\PV\frac{1}{1-x}\int[dk_T]\, e^{i\vec{k}_T\cdot\vec{b}_T}\,\frac{p\,x^2}{k_T^2}\,
\frac{1}{\sqrt{k_T^2+p^2x^2}}, \nonumber \\
R^a_3 &= -\frac{1}{2}\asc\,\PV\frac{1}{p(1-x)}\int[dk_T]\, e^{i\vec{k}_T\cdot\vec{b}_T}\,
\frac{1}{\sqrt{k_T^2+p^2(1-x)^2}}, \nonumber \\
R^a_4 &= \frac{1}{2}\asc\,\PV\frac{1}{p(1-x)}\int[dk_T]\, e^{i\vec{k}_T\cdot\vec{b}_T}\,
\frac{1}{\sqrt{k_T^2+p^2x^2}},
\end{align}
and $\tilde{B}^{(1)}_{\mathrm{sail,r}} = R^a_1+R^a_2+R^a_3+R^a_4$.

\subsection{Vertex diagram}

The vertex diagram has only real contributions, which are the three terms of eq.~\eqref{k0vertex} with $k_3=p(1-x)$,
\begin{align}
R^b_1 &= \frac{1}{2}\asc\,(1-\eps)\,p\int[dk_T]\,e^{i\vec{k}_T\cdot\vec{b}_T}\,
\frac{(1-x)^2}{k_T^2\sqrt{k_T^2+p^2(1-x)^2}}, \nonumber \\
R^b_2 &= \frac{1}{2}\asc\,(1-\eps)\,p\int[dk_T]\,e^{i\vec{k}_T\cdot\vec{b}_T}\,
\frac{x(1-x)}{k_T^2\sqrt{k_T^2+p^2x^2}}, \nonumber \\
R^b_3 &= \frac{1}{2}\asc\,(1-\eps)\,p\int[dk_T]\,e^{i\vec{k}_T\cdot\vec{b}_T}\,
\frac{x}{(k_T^2+p^2x^2)^{3/2}},
\end{align}
and $\tilde{B}^{(1)}_{\mathrm{vertex}} = R^b_1+R^b_2+R^b_3$.

\subsection{Wilson-line diagram}

The virtual contributions of the Wilson-line diagram are
\begin{align}
V^c_1 &= -\frac{1}{2}\asc\,\delta(1-x)\int_{-\infty}^{\infty}dk_3\int[dk_T]\,\frac{1}{E}\,
\frac{2-e^{-ik_3L}}{(k_3+i0)(k_3-i0)}, \nonumber \\
V^c_2 &= -\frac{1}{2}\asc\,\delta(1-x)\int_{-\infty}^{\infty}dk_3\int[dk_T]\,\frac{1}{E}
\left[\frac{i(b^z-L)}{k_3+i0}+\frac{iL}{k_3-i0}\right], \nonumber \\
V^c_3 &= \frac{1}{2}\asc\,\delta(1-x)\int_{-\infty}^{\infty}dk_3\int[dk_T]\,
\frac{1-e^{ik_3L}}{(k_3+i0)(k_3-i0)}\,\frac{e^{i\vec{k}_T\cdot\vec{b}_T}}{E}, \nonumber \\
V^c_T &= \frac{1}{2}\asc\,\delta(1-x)\int_{-\infty}^{\infty}dk_3\int[dk_T]\,\frac{1}{E}
\left[-\frac{1-e^{i\vec{k}_T\cdot\vec{b}_T}}{(k_b-i0)(k_b+i0)}-\frac{i\,b_T}{k_b+i0}\right],
\end{align}
and the real contributions are
\begin{align}
R^c_1 &= \frac{1}{2}\asc\,p\int[dk_T]\,\frac{1}{E}\,\frac{e^{ik_3L}}{k_3^2}\bigg|_{k_3=p(1-x)}, \nonumber \\
R^c_2 &= \frac{1}{2}\asc\,p\int[dk_T]\,\frac{1-e^{ik_3L}}{k_3^2}\,\frac{e^{i\vec{k}_T\cdot\vec{b}_T}}{E}
\bigg|_{k_3=p(1-x)}.
\end{align}
$V^c_1$ is the self-energy of the leg of length $L$, $V^c_3$ and $R^c_2$ are the virtual and real contributions of the exchange
between the legs, $R^c_1$ is the self-energy of the leg of length $L-b^z$, and $V^c_T$ is the self-energy of the transverse
link. The terms linear in the lengths, $V^c_2$ and the last term of $V^c_T$, vanish in dimensional regularization, since with
the length factored out the integrals over $k_3$ and $\vec{k}_T$ contain no scale. Taking the real part, the bracket in $V^c_2$ becomes $-\pi[L+(L-b^z)]\delta(k_3)$, proportional to the total
length of the two legs. The correspondence with the results in section~\ref{secresults} is summarized in table~\ref{tabcontrib}.

\bibliographystyle{JHEP1}
\bibliography{qtmdpdf}

\end{document}